# First-principles Approaches to Simulate Lithiation in Silicon Electrodes

Qianfan Zhang[1,2], Yi Cui[2,3] and Enge Wang[4*]

1 School of Material Science and Engineering, Beihang University, Beijing 100191, China
2 Department of Materials Science and Engineering, Stanford University, Stanford, CA 94305, USA.
3 Stanford Institute for Materials and Energy Sciences, SLAC National Accelerator Laboratory, Menlo Park, CA 94025, USA.
4 School of Physics, Peking University, Beijing 100871, China

*Corresponding author, email: egwang@pku.edu.cn

## Abstract

Silicon is viewed as an excellent electrode material for lithium batteries due to its high lithium storage capacity. Various Si nano-structures, such as Si nanowires, have performed well as lithium battery anodes and have opened up exciting opportunities for the use of Si in energy storage devices. The mechanism of lithium insertion and the interaction between Li and the Si electrode must be understood at the atomic level; this understanding can be achieved by first-principles simulation. Here, first-principles computations of lithiation in silicon electrodes are reviewed. The review focuses on three aspects: the various properties of bulk Li-Si compounds with different Li concentrations, the electronic structure of Si nanowires and Li insertion behavior in Si nanowires, and the dynamic lithiation process at the Li/Si interface. Potential study directions in this research field and difficulties that the field still faces are discussed at the end.

## 1. Introduction

Energy storage is a crucial aspect of integrating renewable energy sources in power grids, which makes the development of efficient high-capacity batteries an important technological field [1]. Li ion batteries have been the most important portable power source for consumer electronics and show great promise for vehicle electrification. Carbon materials, usually in the form of graphite, which has a theoretical capacity of 372 mAh/g, are commonly used as the negative electrodes in commercial rechargeable lithium ion batteries. To meet the energy goals of high capacity and energy density, materials possessing ultra-high lithium capacities are being investigated. Compared with graphite, Si can store ~10 times more Li, with a theoretical gravimetric energy density of 3572 mAh/g [2-4]. Moreover, Si is safer, less expensive, and far more abundant than graphite. However, the electrochemical cycling performance is hampered by massive structural changes and volume expansion on the order of 300% [5-7], resulting in electrode particle fracture, disconnection between the particles, capacity loss, and thus very limited cycle life.

Excitingly, silicon nanowires (SiNWs) have recently been demonstrated as the basis for ultra-high capacity lithium ion battery negative electrodes [8], which opens up amazing opportunities for silicon electrode-based energy storage devices. SiNWs possess excellent properties for Li insertion, such as efficient electron transport along the axis and large Li ion flux due to the high surface area to volume ratio. Most importantly, the success of using SiNWs lies in their facile strain accommodation, which can address the mechanical fracture issues. Beyond SiNWs, a variety of Si nanostructure morphologies has subsequently been shown to overcome the mechanical fracture issues and perform well as anodes, including crystalline amorphous core-shell SiNWs [9], carbon-amorphous Si core-shell NWs [10], Si nanotubes [11], porous Si particles [12], hollow nanoparticles [13], carbon-silicon composites [14, 15] and coated-hollow structures [16-18].

During the lithiation process, the structure, crystalline phase, electronic properties and mechanical properties will undergo significant changes, and these changes critically affect the performance of the lithium battery. Experiments have been performed to address these issues and trace these changes, including X-ray diffraction (XRD) [19], nuclear magnetic resonance (NMR) [20,21] and electron energy loss spectroscopy (EELS) [22]. However, in many cases, experiments can only give properties that are averaged over time or space, while the local and detailed information is buried. Therefore, it is important to study the lithiation process in Si electrodes at the atomic level and quantitatively understand the fundamental interactions between Li and Si atoms in the anode.

Lithiation in Si electrodes can be studied quantitatively with first-principles calculations based on density-functional theory (DFT) [23,24], which is widely used in investigating the electronic structure and studying a wide range of different atomic systems, including molecules, surfaces, nano-structures and bulk materials. In this review paper, we summarize and discuss the theoretical and computational studies on the issue of Li insertion in various kinds of Si materials, such as bulk Si, SiNWs and Si surfaces. This review discusses many aspects in this field and tries to fully introduce the detailed atomistic picture of the lithiation properties and lithiation process in Si electrode that have been obtained through first-principles simulations.

The review is organized in the following manner: In Section 2, fundamental theories and computational *ab initio* schemes for simulating the Li-Si system are briefly introduced. In Section 3, various properties of bulk Li-Si alloys, including their geometry, phase transitions, voltage, and Li diffusivity, are systemically reviewed. Section 4 contains a discussion of the electronic properties of SiNWs and single Li insertion behavior in SiNWs. In Section 5, the lithiation dynamics at different Si surfaces or Li/Si interface systems are summarized. In Section 6, we discuss the potential directions in this field and the advanced computational techniques that can overcome the remaining difficulties. Finally, in Section 7, concluding remarks are given.

## 2. Theoretical Schemes

### 2.1 First-principles Methods

First-principles electronic structure methods, usually performed by using DFT, are quantum mechanical methods for numerically solving the Schrödinger (nonrelativistic) or Dirac (relativistic) equation for systems of electrons. The term "first-principles" simply means that there is no empirical fitting, or equivalently, no adjustable parameters. The term "*ab initio*" is sometimes used instead and means the same thing. By self-consistently solving Kohn-Sham one-electron equations [24], the ground state of a system composed of many electrons and nuclei can be attained [23]. For treating the exchange-correlation energy and many-body potential, the local density approximation (LDA) [25] or the generalized gradient approximation (GGA) [26,27] is introduced. First-principles schemes can provide total energies, stable atomic geometries and various electronic properties of molecular or crystalline materials.

To deal with the Kohn-Sham equation, band calculation schemes are used. Nowadays, the pseudopotential methods are commonly adopted because they only consider the valence electrons that interact with the electrons of other atoms, and can thus significantly reduce the computational cost. The pseudopotentials can be classified into three types: norm-conserving pesudopotentials (NCPP) [28], ultrasoft pseudopotentials (USPP) [29] and projector augmented wave pseudopotentials (PAW) [30,31]. USPP and PAW methods are often applied because of the smoothness of wave functions, which means relatively small numerical arrays or matrices for computation.

For numerical calculations, a plane-wave basis set is often used because it has a high degree of flexibility, provided that a sufficient quantity of waves is used to model fine details in the variation of the electron density. To overcome the huge-size eigenvalue problem, new iterative algorithms such as conjugate gradient [32], RMM-DIIS [33] (residual minimization) and Block-Davidson [34] schemes have been introduced, while combining effective mixing methods. These schemes are already coded in various package programs such as VASP [33], CASTEP [35] and ABINIT [36].

For treating large supercells, other methods beyond plane-wave basis sets are introduced. Real-space grid schemes [37-39], in a similar pseudopotential framework, have been developed. Finite-difference or finite-element algorithms are used to obtain eigen-functions in real space in contrast to the traditional *k*-space methods. This real-space method is more suitable for parallel computation, and results in highly efficient calculations. A scheme using a local-orbital basis set is another powerful one for dealing with large supercells [40,41], where atomic orbital-like functions expressed by Gaussians or numerical functions are used as a basis set of the linear combination of atomic orbitals. These are well known as "order-$N$" algorithms. However, the price that must be paid is that results will be constrained by the flexibility of the basis set, and the incompleteness of the basis set is a serious problem compared with the plane-wave scheme.

**2.2 Schemes for Simulating Lithiation or Delithiation Processes**

Molecular dynamics (MD) can model the detailed microscopic dynamical behavior of many different types of systems found in chemistry, physics or biology, and it is one of the most powerful tools to study the equilibrium and transport properties of many–body systems. For *ab initio* MD [42,43], the simulation is performed based on self-consistent Kohn–Sham energy computation. The basic idea is to compute the forces acting on an ion from electronic structure calculations that are performed on-the-fly as the molecular dynamics trajectory is generated. In this way, the electronic variables are not integrated out beforehand, but are considered as active degrees of freedom, and their evolutions are approximately decoupled with those of ions. This implies that, given a suitable approximate solution of the many-electron problem, "chemically complex" systems can be handled. The real-time nuclear motion of the particles is modeled step-by-step using the laws of classical Newtonian mechanics, the flow of which can be summarized as below:

1. Compute the total energy for the simulated system and the force on each ion.
2. Obtain the coordinate and velocity for each ion at the following time step.
3. Return to step 1 with the updated coordinates and velocities.

For many simulation cases, the molecular dynamics of lithiation using DFT is too computationally costly. V. L. Chevrier *et al.* developed the following protocol to simulate the lithiation of Si at room temperature while avoiding the lengthy diffusion dynamics [44,45]. The lithiation process was modeled using the following sequence:

1. Add a Li atom at the center of the largest spherical void.
2. Increase the volume and scale coordinates.
3. Allow coordinate optimization at a fixed volume.
4. Calculate the total energy.
5. Return to step 1 with the current structure until the desired x in $Li_xSi$ is reached.

Correspondingly, the delithiation process was modeled as follows:

1. Randomly remove a Li atom from the system.
2. Decrease the volume and scale coordinates.
3. Allow coordinate optimization at a fixed volume.
4. Calculate the total energy.
5. Return to step 1 with the current structure until the desired x in $Li_xSi$ is reached.

In this protocol, a new Li atom is inserted into the largest void within the computational cell at each step. However, M. K. Y. Chan *et al.* have tested the insertion of Li into different interstitial sites, and found that in only approximately 20% of all cases, the site farthest away from existing atoms is the most energetically favorable site after relaxation [46]. Therefore, they modify the algorithm above to allow the insertion of Li into all possible interstitial sites, and the lowest energy configuration after relaxation is chosen at each step. The details for each lithium atom insertion are described as below:

1. Using a regular spatial grid, all interstitial sites { $r_i$ } that are at least $d_{min}$ from the nearest atoms are determined.

2. A lithium atom is inserted separately into each $r_i$, and the total energy is calculated.

3. The relaxed configuration with the lowest energy is relaxed and chosen for further lithium insertion.

The value of $d_{min}$ is set to the largest value such that at least one interstitial site is found at all steps, which is suggested to be 1.9 Å.

## 2.3 Computation of Physical Quantities

### 2.3.1 Energy

The total energy of a simulated Li-Si compound can be directly calculated after the DFT self-consistent simulation. Based on this, the formation energy (or binding energy) can be calculated, which is very important in determining the structure and judging the stability of Li-Si compounds. The definitions of formation energy are somewhat different in different papers, but they all have similar forms. The general expression can be given as

$$E_f = E_{Li_xM} - xE_{Li} - E_M \tag{1}$$

Where $E_M$ and $E_{Li_xM}$ represent the energies for the pure Si structure before lithiation and the Li-Si alloy after lithiation, respectively. $E_{Li}$ represents the energy of Li in Li metal, or sometimes the energy of an isolated Li atom.

In some papers, a similar parameter called "mixing enthalpy per atom ($\Delta E_{mix}$)" is used for the energy comparison between different concentrated Li-Si alloys [47]. It can be expressed as

$$\Delta E_{mix} = E_{Li_xSi_{1-x}} - xE_{Li} - (1-x)E_{Si} \tag{2}$$

where $E_{Li_xSi_{1-x}}$ is the total energy per atom of the Li-Si alloy in question, x is the atomic fraction of Li, and $E_{Li}$ and $E_{Si}$ are the total energies per atom of c-Si and bcc-Li, respectively.

### 2.3.2 Voltage

For a two-phase reaction between Li$_x$M and Li$_y$M, the voltage relative to Li is given by the negative of the reaction free energy per Li [48], that is

$$V = -\left\{ \frac{G_{Li_yM} - G_{Li_xM}}{y - x} - G(Li_{metal}) \right\} \tag{3}$$

where G is the Gibbs free energy

$$G = E + pV - TS \tag{4}$$

and $Li_{metal}$ refers to the Li atom in metal. The enthalpic ($pV$) contribution to $G$ is of the order ~10 µeV per Li at atmospheric pressure. The entropic ($TS$) contribution to the voltage is estimated to be systematic and of order ~10 meV. Therefore, it is safe to replace $G$ in Equation 4 with total energies ($E$) from DFT calculations. If the total energies over the whole x range of Li$_x$Si compounds are calculated by first-principles method, the voltage curve for lithiation process can be achieved.

### 2.3.3 Mechanical properties

Elastic constants, $C_{ij}$, can be obtained by computing the energies of deformed unit cells; the deformation strain tensor, $e$, with six independent components is represented as

$$e = \begin{pmatrix} e_1 & e_6/2 & e_5/2 \\ e_6/2 & e_2 & e_4/2 \\ e_5/2 & e_4/2 & e_3 \end{pmatrix} \tag{5}$$

For cubic phases, orthorhombic, isotropic, and monoclinic distortions were applied to obtain the three independent elastic constants, $C_{11}$ $C_{12}$ and $C_{44}$ (expressed using Voigt notations). For tetragonal phases, six independent deformation modes were applied to calculate $C_{11}, C_{12}, C_{13}, C_{33}, C_{44}$ and $C_{66}$. Self-consistent relaxation is allowed in all strained unit cells, and the total energy change with respect to the strain tensor gives [49,50]

$$E(e) = E(0) - p(V)\Delta V + \frac{V}{2}\sum_{ij} C_{ij} e_i e_j + O[e^3] \qquad (6)$$

where $E(0)$ and $E(e)$ are the internal energies of the initial and strained lattices, respectively; $V$ is the volume of the unstrained lattice; $p(V)$ is the pressure of the undistorted lattice at volume $V$; $\Delta V$ is the change in the volume of the lattice due to the strain; $e$ is the strain tensor; and the $O[e^3]$ term can be neglected.

Once $C_{ij}$ values are known, mechanical quantities, such as bulk, shear and Young's moduli, symbolized as $B$, $E$ and $G$, can be calculated using the following expressions:

$$B = \frac{1}{9}(C_{11} + C_{22} + C_{33} + 2C_{12} + 2C_{13} + 2C_{23}) \qquad (7)$$

$$G = \frac{3(C_{11} - B)}{4} \qquad (8)$$

$$E = \frac{9B(C_{11} - B)}{C_{11} + 3B} \qquad (9)$$

### 2.3.4 Diffusivity

To better understand the dynamic properties, diffusion constants of Si and Li can be calculated by performing *ab initio* MD simulations. First, average mean-square displacements (MSD) of Li or Si atom as a function of MD time steps for different temperatures can be calculated by sampling configurations. MSD can be expressed as $\langle |R_i(t) - R_i(0)|^2 \rangle$, $R_i$ is the atomic position broken brackets denote thermal averages, and $t$ is the time. The diffusion constants $D$ of Si and Li were calculated based on the Einstein relation

$$D = \frac{\langle |R_i(t) - R_i(0)|^2 \rangle}{6t} \qquad (10)$$

## 3. Bulk Li-Si Compounds

## 3.1 Single Li Atom in Bulk Si

We start our review by discussing single Li atom insertion in crystalline Si, because it is the basis for understanding the mechanism of the Li insertion process and the interaction between Li and the Si host. W. H. Wan *et al.* reported a detailed study on this topic [51]. For simulation, a single Li atom was put into a supercell with 64 Si atoms. Various nonequivalent insertion positions in the Si lattice were examined and their binding energies were calculated to extract the energetically favorable sites. The positions and corresponding values of binding energies are shown in Figure 1 (a) and (b) respectively, including the tetrahedral ($T_d$), hexagonal (Hex), anti-bonding (A), bond-center (B), center of the second nearest silicon (C), and midway (M) sites between the Hex site and B site. The Td site is the most stable position, on which there are four nearest-neighbor Si atoms with Li–Si distances of 2.45 Å. Then, based on the energy calculation of different Li doping sites, the diffusion barrier can be achieved using the NEB scheme by relaxing various configurations between the starting and ending points along the diffusion pathway [52,53]. The energy curve for Li diffusion between the adjacent Td sites is presented in Figure 1 (c). The barrier height is about 0.58 eV, which is in good agreement with experiment (0.57–0.79 eV) [54]. The Hex site, on which there are six nearest-neighbor Si atoms with Li–Si distances of 2.37 Å, is the transition state along the migration pathway. The long-range diffusion processes occurs in a manner that can be called a 'zig-zag' fashion.

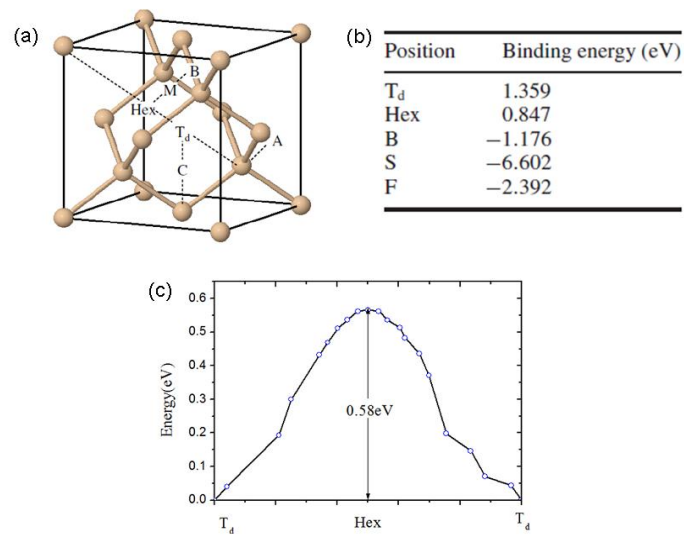

Figure 1. (a) The $T_d$, Hex ,A, B, C, and M sites in bulk Si with a diamond structure. (b) The binding energies of single Li dopants in bulk Si. (c) Diffusion energy for a Li dopant along the $T_d$–Hex–$T_d$ diffusion pathway. The figure is from Ref. [51].

There are also some other simulation works on this topic, and it has been commonly accepted that the $T_d$ site is the insertion position for Li atoms and the $T_d$-Hex-$T_d$ trajectory is the diffusion pathway; other calculated diffusion barriers are similar to the one shown above, e.g. 0.60 eV by H. Kim *et al.* [55] or 0.55 eV by K. J. Zhao *et al.* [56].

## 3.2 Geometry Evolution During Lithiation Process

As more and more Li atoms are inserted into bulk Si to form $Li_xSi$ compounds, the structure undergoes great changes. Experimental studies have provided strong evidence for the formation of various stable Li-Si crystalline phases during high-temperature lithiation (~ 415 ℃), such as $Li_{12}Si_7$, $Li_7Si_3$, $Li_{13}Si_4$, $Li_{15}Si_4$, and $Li_{22}Si_5$ [57]. Many simulations have studied the structures of these crystalline phases [58-62]. Here, we use the data in the paper by V. L. Chevrier *et al.* to show the crystalline configurations of different Li-Si compounds that can be observed in experiment [62]. The upper left panel of Figure 2 lists the number of atoms in the primitive cell, the space group of these phases and the underlying crystalline lattice, while the upper right panel of Figure 2 shows the primitive cells of the binary phases and the conventional cells of Si and Li.

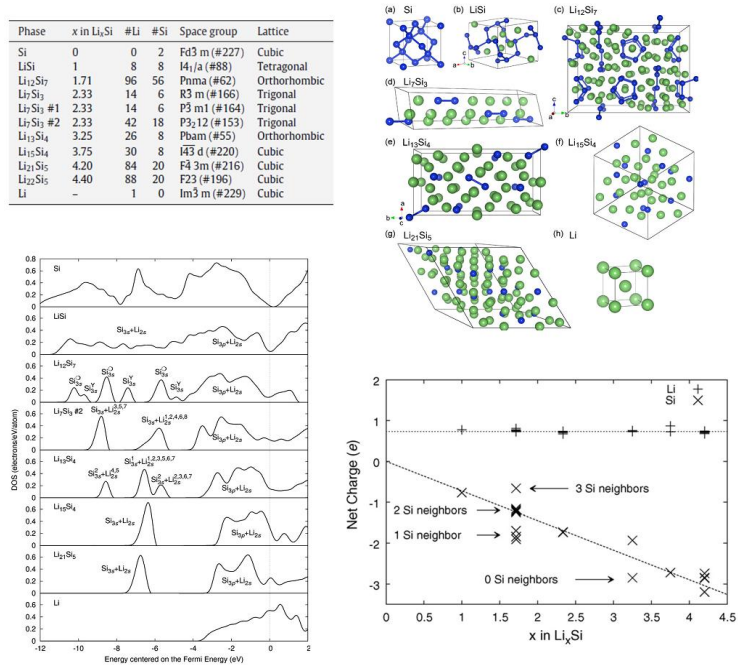

**Figure 2.** Upper left panel: Li, Si and crystalline Li–Si phases. Upper right panel: The structures and the primitive cells of the mixed crystalline Li-Si phases are shown. Partial occupancies in $Li_7Si_3$ are indicated by "pie chart" Li atoms. Left lower panel: total electronic density of states for Si, Li, and the Li–Si phases.

**Peaks are labeled according to the principal orbital contribution. Right lower panel: The net charge in e of the Li and Si atoms in the crystalline Li–Si phases: LiSi, $Li_{12}Si_7$, $Li_7Si_3$ #2, $Li_{13}Si_4$, $Li_{15}Si_4$, $Li_{21}Si_5$. The figure is from Ref [62].**

In contrast to high-temperature lithiation, room-temperature Si lithiation frequently leads to amorphous lithium silicide (a-$Li_x$Si), which is a particularly important phase transition and structural change [63]. Using *ab initio* schemes, the amorphization process and structural evolution of $Li_x$Si compounds can be studied, and many works have focused on this issue.

W. H. Wan *et al.* explored the phase transition and structural distortion issue by inserting Li atoms into Si one by one and simulated the configuration evolution of $Li_x$Si alloys at $x$ = 0.03125, 0.0625, 0.125, 0.1875 and 0.25 [51]. It is found that when $x < 0.0625$, the most stable configuration is that both Li atoms are located on the $T_d$ sites. The further the distance between them, the lower the energy of the structure, and the crystalline Si framework remains. A clear local structure distortion occurs at $x = 0.125$. The obtained stable structure causes a local distortion of the host lattice, while the homogeneous arrangement of the six-member ring in perfect crystalline Si is destroyed into one five-member ring and one seven-member ring around the Li dopants. As the doping concentration increases to 0.1875 or above, the Si lattice is significantly damaged.

S. C. Jung *et al.* computationally determined the Li/Si ratio at which the crystalline-to-amorphous phase transition occurs, based on the comparison of energies between the crystalline and amorphous phases [64]. Figure 2(a) shows the formation energies of the crystalline and amorphous $Li_x$Si structures. At x = 0, the crystalline phase is more stable by 0.29 eV per Si atom than the amorphous phase. As x increases, the crystalline phase becomes less stable and the amorphous phase becomes more stable, resulting in the crystalline-to-amorphous phase transition. According to formation energy curves, the phase transition occurs at about x = 0.3, which is very low compared with the ratio of x = 3.75 in the fully lithiated $Li_{15}Si_4$ phase, and indicates that the crystalline-to-amorphous transition of the silicon electrode occurs during the initial discharge process. This calculated ratio of 0.3 Li atoms per Si atom is in good agreement with the 0.26:1 ratio from the $^7$Li NMR spectra, at which point the new $^7$Li peaks at 105 mV appears, implying a breaking of the crystalline silicon [20].

H. Kim *et al.* compared the mixing enthalpies (see Section 2.3 for reference) of amorphous and crystalline Li-Si alloys with a relatively high Li concentration [47]. The results are shown as Figure 3(b). For amorphous phases, the value of $\Delta E_{mix}$ drops as the Li concentration increases, and changes from positive to negative at 40 atom% Li. Above this Li concentration, the mixing enthalpy continues to decrease and falls to a valley plateau between 60 and 80 atom % Li, with an energy gain of 0.16-0.18 eV/atom with respect to c-Si and bcc-Li. The calculations are consistent with experiments that the formation of a-$Li_{2.1}$Si (68 atom% Li) can gain 0.12 eV per Si [63]. For crystalline phases, a distinct mixing enthalpy minimum is found at 71 atom % Li, and on average the total energies are ~0.1 eV/atom lower than their amorphous counterparts. Hence, a-Li-Si alloys may undergo recrystallization at elevated temperatures, as evidenced by earlier experiments [3].

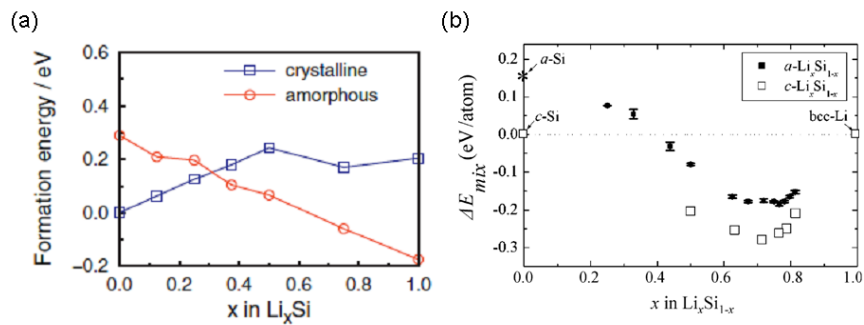

**Figure 3. (a) Formation energies of crystalline and amorphous $Li_xSi$ phases at T = 0 K. (b) Variations in the mixing enthalpies of amorphous and crystalline Li-Si alloys as a function of Li concentration. (a) and (b) are from Ref. 64 and Ref. 47, respectively.**

## 3.3 Electronic Structure Analysis

People often use the electronic density of states (EDOS) to characterize the electronic properties of materials. The lower left panel of Figure 2 shows the EDOS of Si, Li, and the Li-Si crystalline phases simulated by V. L. Chevrier *et al.* [62]. The Fermi energy is marked by a dashed vertical line. The peaks for the EDOS are labeled by atomic orbital components, based on projected EDOS. For Si, the occupations have some s and p character, while for Li, the occupations only have s character. For the energy region near the Fermi level, the states are mainly a hybridized $Si_{3p}$ orbital and $Li_{2s}$ orbital. It is somewhat unexpected to see that the EDOS at the Fermi level for various

Li-Si phases is extremely low even though numerous Li atoms are present in these phases. This phenomenon can be attributed to the Zintl-like nature of the Li-Si alloy [65]. Indeed, Zintl-type phases are often semiconductors, and experimental measurements have demonstrated that crystalline $Li_{12}Si_7$ and $Li_7Si_3$ own semiconductor-like features [66,67].

Using Bader charge analysis [68], V. L. Chevrier *et al.* also calculated the net charge of the Li and Si atoms as a function of x in c-$Li_x$Si, as shown in the lower right panel of Figure 2 [62]. In all Li-Si phases, the Li atoms have a similar positive charge of 0.73e, shown by the dotted line. As the Li concentration increases, the net charge of the Si atoms becomes more negative, and there is clearly local-environment dependent behavior, which varies according to the number of Si neighbors. For instance, in the $Li_{12}Si_7$ structure, there are Si atoms with three, two, and one Si neighbors, which lead to clearly separated atoms in terms of charge; the Si atoms with one Si neighbor have similar charges in $Li_{12}Si_7$, $Li_7Si_3$, and $Li_{13}Si_4$, and the isolated Si atoms of phases $Li_{13}Si_4$, $Li_{15}Si_4$, and $Li_{21}Si_5$ also have similar charges. This charge transfer analysis confirmed the traditional understanding that the Li–Si alloys are Zintl-like phases where the amount of charge transferred to the Si atoms depends on the number of Si neighbors, but the Li atoms donate roughly the same amount of electrons in all Li–Si crystalline phases with a very small dependence on proximity to the receiving Si atoms. In another work, Chevrier *et al.* calculated the charge transfer in a-$Li_x$Si. They found similar behavior to the c-$Li_x$Si case, but a slightly lower value of 0.68 electrons per effective Li neighbor was yielded [45].

### 3.4 Voltage Computation

Voltage is one of the most important parameters for lithiation experiments, and it is also the essential point to evaluate the performance of a battery. Therefore, it is necessary to compute the evolution of voltage during the whole lithiation process, from the fully-unlithiated to the fully-lithiated stage.

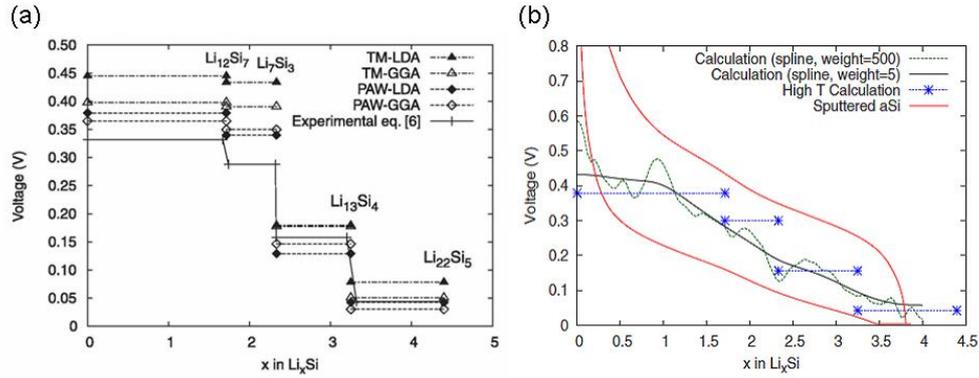

**Figure 4.** (a) Experimental and calculated voltage-composition curves of a Li/Li$_x$Si electrochemical cell at high temperature (415 ℃). (b) Voltage–composition curves of silicon lithiation calculated from the lithiation protocol by using smoothing weights of 5 and 500 (see Ref [44] for details) and by using the crystalline structures found at high temperatures, as indicated in the legend. The solid gray red shows the experimental curve obtained from the lithiation of sputtered amorphous silicon. (a) and (b) are from Ref. [69] and Ref. [44], respectively.

V. L. Chevrier *et al.* computed the voltage curve during evolution of crystalline Li-Si phases at high temperature (415 ℃). Figure 4(a) shows the simulated and experimental results, and different dashed lines are obtained by different pseudopotential schemes [69]. The calculation of the voltage curve is based on the formation energies of crystalline phases arising during lithiation – Li$_{12}$Si$_7$, Li$_7$Si$_3$, Li$_{13}$Si$_4$ and Li$_{22}$Si$_5$, and four voltage plateaus are established. From the comparison, it can be seen that the simulation can correctly reproduce the shape of the experimental curve [57].

V. L. Chevrier *et al.* also simulated the voltage evolution of the lithiation in amorphous Si at room temperature, using the lithiation protocol they developed (see Section 2 for reference) [44]. Figure 4(b) shows the voltage–composition curves obtained from the smoothing spline, from the experimental sputtered a-Si, and from calculations using the crystalline phases of Li–Si. The smoothing spline is obtained by taking the average of the formation energy values at every Li concentration and by constructing a curve piecewise from cubic polynomials. The agreement with experiment confirms that the essential physics of silicon lithiation is correctly captured by the simulation. They also apply protocol variations for comparative study, which shows that their simulation scheme is reasonable. Beyond this work, they do further research on similar "disordered lithiated silicon" using the same algorithm as above, and also add simulations of the delithiation process [45].

## 3.5 Mechanical properties

Mechanical stability is one of the key criteria for the selection of battery materials [69], and a successful electrode material has to maintain its mechanical integrity and chemical properties over a long cycling lifetime. Therefore, it is necessary to study the mechanical properties of Li-Si compounds from the point of view of theoretical simulations, and how to overcome the pulverization and improve the mechanical behavior has become an important topic for computational researchers.

H. Kim *et al.* investigated the variations in volume and density as a function of Li content [47]. The dependence of volume on Li concentration for c-Si and a-Si are shown in Figure 5. Here, the volume of each alloy is normalized with respect to that of c-Si (in which each Si occupies a volume of $\approx 20.47$ Å$^3$). For both crystalline and amorphous phases, the volume increases nearly linearly with Li content, while the opposite trend is observed for the density values. As expected, the crystalline phase is slightly denser than the amorphous alloy of corresponding composition. The fully lithiated a-Li$_{4.33}$Si (c-Li$_{4.4}$Si) phase is predicted to yield a 334 (296) % volume expansion, which is in good agreement with $\approx 300\%$ from previous experimental measurements [70]. Many other works also studied the volume expansion and similar results are shown [44,55,56].

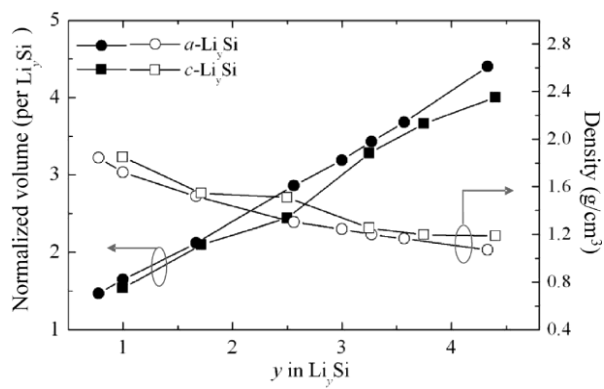

**Figure 5. Variation in the volume and density of amorphous and crystalline Li-Si alloys as a function of Li concentration, as indicated. The volume (per Li$_y$Si) of each alloy is normalized with respect to that of c-Si. The figure is from Ref. [47].**

V. B. Shenoy *et al.* reported a systemic study on the mechanical properties of lithiated c-Si and

a-Si [71]. Figure 6 shows the calculated elastic constants $C_{ij}$ for c-Li-Si alloys, and the variation of the bulk modulus $B$, shear modulus $G$, Young's modulus $E$ and Poisson's ratio $\vartheta$ for both c-Li$_x$Si and a-Li$_x$Si. One can see that for the crystalline systems, the bulk, shear, and Young's moduli depend very strongly on Li concentration, which decrease linearly and show significant softening in the Li-rich phases. The moduli for the most highly-lithiated phases, Li$_{15}$Si$_4$ and Li$_{22}$Si$_4$, are nearly an order of magnitude smaller than the corresponding values for Si. This phenomenon can be attributed to the decrease in the population of strong covalent Si–Si bonds, and the increase in the population of much weaker ionic Li–Si bonds. From the plots of the elastic moduli in Figure 6, it is clear that the slopes of the moduli for amorphous systems are smaller than that of crystalline phases, which indicates a weaker softening effect. This can be understood by noting that the moduli of pristine amorphous Si are smaller than those of cubic Si by about 30–50%, while the moduli of amorphous Li and bcc Li are nearly equal.

| Phase | $C_{11}$ | $C_{22}$ | $C_{33}$ | $C_{12}$ | $C_{13}$ | $C_{23}$ | $C_{44}$ | $C_{55}$ | $C_{66}$ |
|---|---|---|---|---|---|---|---|---|---|
| Si | 152.168 | 152.168 | 152.168 | 56.842 | 56.842 | 56.842 | 75.034 | 75.034 | 75.034 |
| LiSi | 101.160 | 101.160 | 74.506 | 20.645 | 37.343 | 37.343 | 55.435 | 55.435 | 36.000 |
| Li$_{12}$Si$_7$ | 92.500 | 95.905 | 88.837 | 4.987 | 11.403 | 8.801 | 30.168 | 34.681 | 44.939 |
| Li$_{13}$Si$_4$ | 77.207 | 64.702 | 82.247 | 18.503 | 7.412 | 6.746 | 73.637 | 32.866 | 40.595 |
| Li$_{15}$Si$_4$ | 46.644 | 46.644 | 46.644 | 21.907 | 21.907 | 21.907 | 28.019 | 28.019 | 28.019 |
| Li$_{22}$Si$_5$ | 89.847 | 89.847 | 89.847 | 1.740 | 1.740 | 1.740 | 30.954 | 30.954 | 30.954 |
| Li | 17.952 | 17.952 | 17.952 | 8.936 | 8.936 | 8.936 | 12.423 | 12.423 | 12.423 |

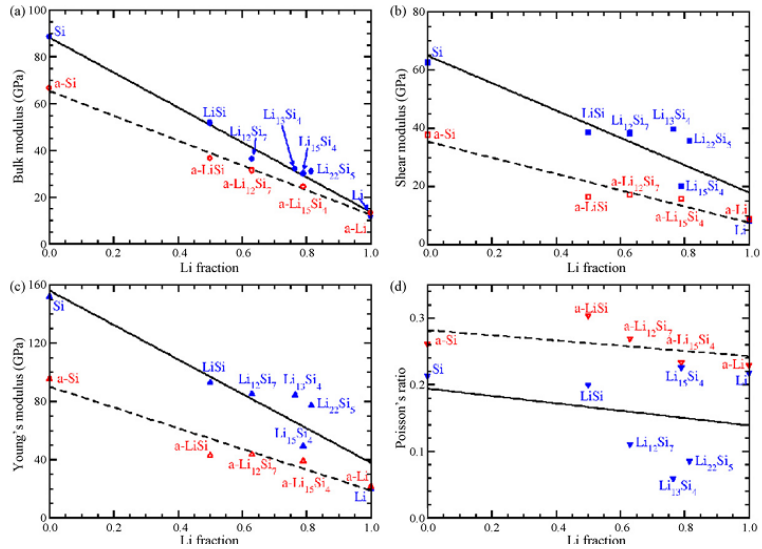

**Figure 6.** Upper panel: Calculated elastic constants $C_{ij}$ for Li, Si and crystalline Li–Si alloys. All quantities are in units of GPa. Lower panel: (a) Bulk modulus $B$, (b) shear modulus $G$, (c) Young's modulus $E$, and (d) Poisson's ratio $\vartheta$ of Li–Si alloys plotted as a function of Li fraction in crystalline (solid symbols) and in amorphous (open symbols) phases for the alloy Li$_x$Si. The Li fraction in the alloy is given by $x/(1+x)$. Solid and broken lines show linear fits for the crystalline and the amorphous systems, respectively. The figure is from Ref. [71].

Many experiments indicate that the large deformation of Si electrodes during lithiation can be accommodated by plastic flow: during lithiation, silicon films deform plastically when the stress exceeds a yield strength [72]. This issue is especially relevant in nanostructured electrodes with confined geometries. Motivated by this phenomenon, K. J. Zhao *et al.* elucidated the plastic deformation in lithiated silicon under uniaxial tension [56]. At the initial stage, a supercell containing 64 atoms was prepared as the a-Si model system, and different amounts of Li atoms, with the largest concentration $f = 0.5$ (f represents the ratio of the number of Li atoms over the number of Si atoms), were then gradually added. To simulate tension, a stress level along the x direction of the structure was prescribed, and then the nominal strain after full relaxation was measured. The stress-strain response curves are shown in Figure 7(a). The solid symbol curves represent the loading paths, while the open symbol curves represent the unloading paths. It is evident that a brittle-to-ductile transition occurs as the lithium concentration increases because of the different behavior in the low Li concentration and high Li concentration cases. In the pure silicon structure, loading leads to nonlinear elastic behavior and the unloading path follows the loading path exactly, which suggests no permanent deformation after unloading. In the case of lithium concentration $f = 0.125$, a small permanent strain of $\varepsilon = 1.21\%$ is observed after unloading. As the Li concentration increases to $f = 0.25$ and $f = 0.50$, the stress-strain curves show substantial plastic deformation. The network can be stretched by 33.5% and 40.5%, respectively, without fracture, and after unloading, large permanent deformation remains in both cases. Figure 7(b) shows the average value of Si coordination, $\langle C_{Si} \rangle$, as a function of applied strain.

The results strongly indicate that the average coordination is closely related to the deformation behavior of lithiated silicon. At lithium concentrations $f = 0$ and $f = 0.125$, the coordination of silicon changes little during deformation, corresponding to brittle behavior at low Li concentration. In contrast, at larger Li concentrations $f = 0.25$ and $f = 0.50$, the Si coordination decreases dramatically with strain, which is related to the ductile behavior.

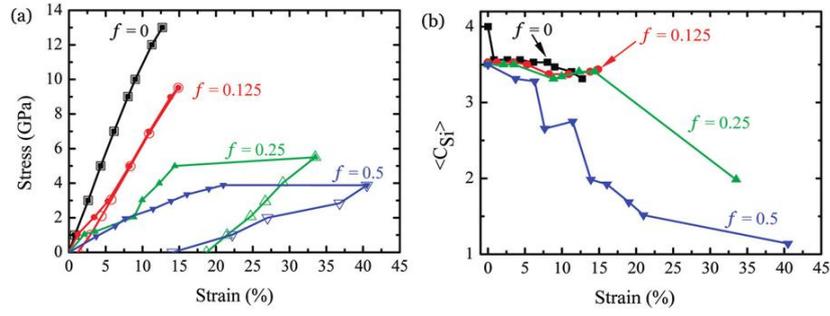

**Figure 7. (a) The stress-strain response of lithiated a-Si under uniaxial tension. The solid symbol lines represent the loading path, while the open symbol lines represent the unloading path. (b) The average value of Si-Si coordination ($\langle C_{Si} \rangle$) as a function of applied strain during loading. The figure is from Ref. [56].**

### 3.6 Diffusion behavior

The diffusion coefficient is the common physical quantity to characterize mobility. For Li-Si compounds, this coefficient can be computed by *ab initio* molecular dynamics methods. Until now, such studies are rare and have usually concentrated on high temperature conditions, which can lead to high mobility and makes the simulation easier and more accurate.

H. Kim *et al.* performed *ab initio* molecular dynamics (MD) simulations in the canonical ensemble to estimate Li and Si mobilities in molten LiSi, $Li_{1.67}Si$, and $Li_{3.57}Si$ alloys at 1050 K [47]. The MD simulation duration is as long as 6 ps, which is sufficient to obtain well-converged results. For LiSi, the diffusion coefficients were predicted to be $D_{Li} = (0.45 \pm 0.04) \times 10^{-4}$ and $D_{Si} = (0.20 \pm 0.05) \times 10^{-4}$ cm$^2$/s; For $Li_{1.67}Si$, $D_{Li} = (0.50 \pm 0.07) \times 10^{-4}$ and $D_{Si} = (0.23 \pm 0.07) \times 10^{-4}$ cm$^2$/s, which are comparable to the values reported by G. A. de Wijs *et al.* [73], $D_{Li} = 0.94 \times 10^{-4}$ and $D_{Si} = 0.42 \times 10^{-4}$ cm$^2$/s (within the simulation time of 2.8 ps); For $Li_{3.57}Si$, the diffusivities increase to $D_{Li} = (0.73 \pm 0.06) \times 10^{-4}$ and $D_{Si} = (0.33 \pm 0.05) \times 10^{-4}$ cm$^2$/s. It can be seen that despite the change in alloy composition, the diffusion coefficient ratio between Li and Si remains more or less constant, $D_{Li}/D_{Si} \approx 2$. This can be expected because of the mass dependence of diffusivity in a liquid-like phase, i.e., $D_2/D_1 \sim (m_1/m_2)^{1/2}$ for a disparate-mass binary mixture. Given that the atomic masses for Li and Si are 6.94 and 28.09 amu, respectively, the calculation result is consistent with what would be expected from the mass dependency, i.e., $D_{Li}/D_{Si} \approx (28.09/6.94)^{1/2} \approx 2.0$.

## 4. Lithium Insertion in Si nanowires

### 4.1 Electronic Properties of Si nanowires

Silicon nanowires (SiNWs) remain one of the most important nano-structured materials since being successfully synthesized about a decade ago [74-76]. SiNWs have become promising for electronic applications in a wide range of areas such as field effect transistors [77-79], nanosensors [80-82] and solar cells [83-85]. Due to the unique properties, such as large surface-to-volume ratio, high carrier transport mobility, and tunable band structure, SiNWs have attracted great interest for their use as a lithium battery electrode. The experimental successes have resulted in extensive theoretical investigations on SiNWs in recent years, and such calculations provide valuable information for understanding their material properties at the nanoscale.

The electronic band structures of SiNWs have been computed using *ab initio* or tight-binding (TB) schemes [86-90]. Studies have found that the band structure varies according to the category and size of SiNWs. Q. F. Zhang *et al.* systematically simulated the band structures and calculated the band gaps of [110], [001], [111], and [112] ultra-thin SiNWs with wire diameters $d < 2.5$ nm [91]. [110], [001], and [111] SiNWs are direct band gap semiconductors, while the series of [112] SiNWs are indirect semiconductors. The calculated band gaps are shown as the left panel of Figure 8. The band gap decreases when the diameter increases. This result agrees well with the commonly belief that significant quantum confinement effects in ultrathin SiNWs can induce enlargement of the band gap. As the size of a SiNW grows further, the essential features of the electronic structure can change. M. F. Ng *et al.* evaluated the diameter-dependent electronic structure in [110] and [100] SiNWs, and their simulation covers the range of diameters up to 7.3 nm, as shown in the right panel of Figure 8 [92]. They also demonstrated the decrease of band gap, and in particular, they found that the direct band gap feature of both ultra-thin [110] and [100] SiNWs starts to change when the diameter increases beyond ~ 4 nm, where there is only a small difference in the direct and indirect band gaps within the experimental measurement uncertainty of ~ 0.1 eV. Therefore, they suggested this is the critical diameter for the start of the gap nature transition for SiNWs. This simulation result is in good agreement with photoluminescence measurements [93].

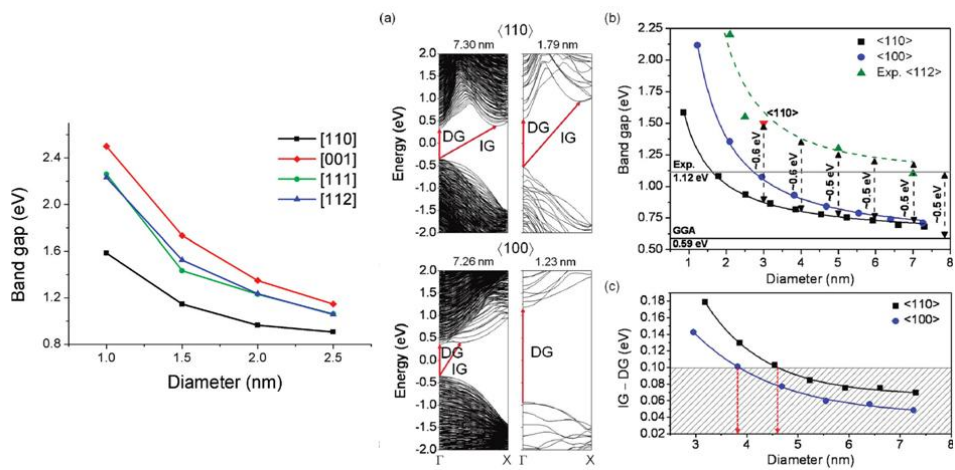

**Figure 8.** Left panel: Band gaps of [110], [001], [111], and [112] SiNWs with wire diameters $d \approx$ 1.0, 1.5, 2.0, and 2.5 nm. Right panel: (a) Band structures of [110] and [100] SiNWs with diameters of about 1 and 7 nm. Arrows indicate the band gaps: DG = direct gap and IG = indirect gap. (b) A plot of band gap as a function of diameter for [110] and [100] SiNWs. Experimental data are fitted with a green dotted line. (c) A plot of IG-DG as a function of diameter for [110] and [100] SiNWs. The filled region indicates <0.1 eV gap difference. Red arrows mark the gap transition size for [110] and [100] SiNWs. Left and right panels are from Ref. [91] and Ref. [92], respectively.

Surface effects are another important issue for the study of SiNWs, because the existence of the surface region can introduce surface electronic states which have different properties than bulk states. On the other hand, the interaction between SiNW surfaces and the surrounding environment is crucial for the potential use of SiNWs in devices. Many factors can influence the characteristics of SiNW surfaces, such as facet features, surface reconstruction and passivation elements. C. R. Leao *et al.* studied the relative contribution of surface Si atoms to the band-edge states, which varies according to the way these surface atoms are bonded to the core atoms [94]. Five categories of [110]-oriented SiNWs, with different diameter and different kinds of facets, are extracted as the simulation models. Figure 9(b) lists the ratios between the Si surface and core atom partial density of edge states – valence-band maximum (VBM) and conduction-band minimum (CBM). They found that Si atoms with different facets have distinct component contributions to the edge states, and generally speaking, the surface atoms have more influence on the VBM state than the CBM state. This indicates that it might be easier to affect the electronic properties of SiNWs with p-type rather than n-type dopants, which has been experimentally observed [95]. X. Xu *et al.* investigated the effects of surface reconstruction and progressive hydroxylation on the electronic properties of [110] hexagonal SiNWs, including band gap, effective mass, and density of states [96]. By increasing the amount of hydroxyl groups on the surface, the size of the band gap decreases. The effective masses for electrons and holes do not significantly change upon hydroxylation, but the reconstruction induces a small decrease in the effective mass of electrons and a small increase in the effective mass of holes.

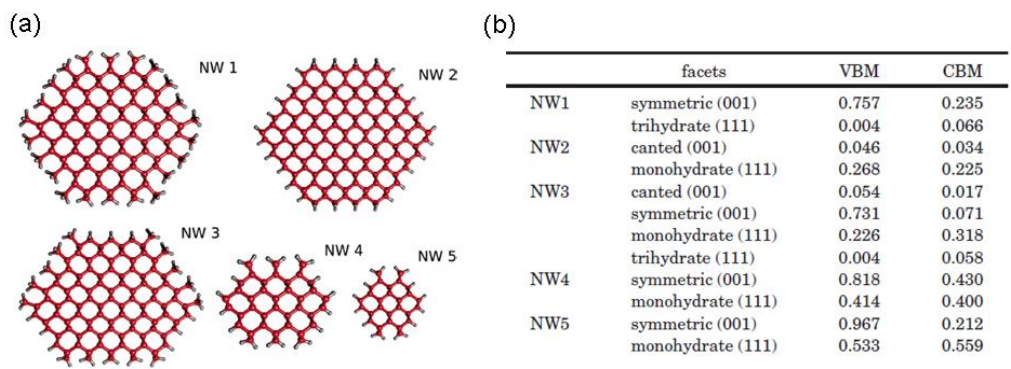

**Figure 9.** (a) Cross-sections of different types of SiNWs studied here. All of them have their axis along the [110] orientation. They differ in the way their facets are terminated or in their relative aspect ratios. The red spheres represent Si atoms and the white ones represent H atoms. (b) Ratio between the Si surface and core atom partial density of the band-edge states. The figure is from Ref [94].

Due to the semiconducting nature of SiNWs, the n-type or p-type doping effect in SiNWs is also interesting for the potential applications in designing high-performance electronic devices. Such doping can be influenced by both quantum confinement and surface effects. J. X. Han *et al.* investigated P-doped Si[110] nanowires by employing a real-space pseudopotential method (see Section 2.1 for reference), and focused on wavefunction, donor ionization energy of the P impurity [97]. Figure 10 shows the spatial distribution of defect charge densities and donor ionization energies in SiNWs with different diameters. The decay of the charge density along the radial coordinate $r$ (Figure 10(c)) and the increase in effective Bohr radius (Figure 10(d)) indicates that as the nanowire diameter decreases, the charge density becomes more localized at the defect center. The donor ionization energy (Figure 10(e)) increases with decreasing diameter. Both of the strong size-dependent phenomena can be attributed to the quantum confinement effect. C. R. Leao *et al.* studied the influence of surface effects on B impurity doping in SiNWs [98]. They found that different substitutional sites will lead to different formation energies for B defects, and such a distinction is large in ultrathin SiNWs but tends to quickly decrease as the diameter of the wires approaches realistic dimensions (more than 30 Å), which indicates that B impurities in SiNWs will be rather uniformly distributed within the wires and on their surfaces beyond the ultra-thin range.

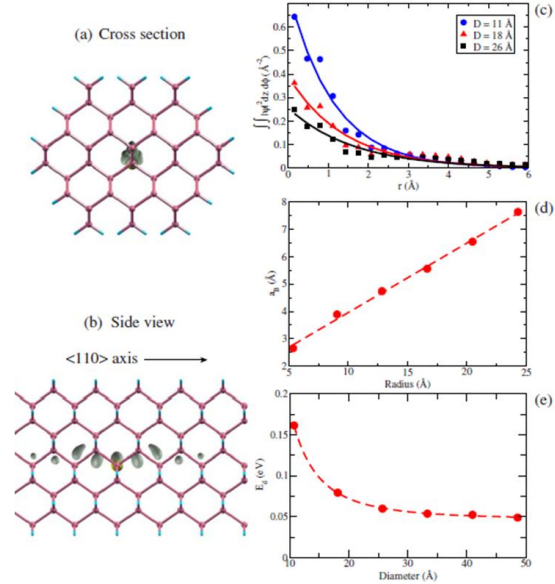

**Figure 10. (a) The cross section and (b) side view of the defect charge density of a P-doped [110] SiNW. The large atom at the center is the P dopant. (c) The defect charge density plotted along the radial direction $r$ with three different diameters $D$ =11, 18 and 26 Å. The curves are fitted using an exponential function with the form $K\exp(-2r/a_B)$. (d) Effective Bohr radius $a_B$ plotted as a function of SiNW radius. The dashed line is a linear fit to our calculated data points: $1.4 + R/3.9$. (e) Donor ionization energy $E_d$ as a function of SiNW diameter. The dashed line is a fit to our calculated data points using a power law. The figure is from Ref [97].**

Beyond the works cited above, there is also some research that paid attention to other interesting properties of SiNWs. For instance, Z. G. Wu *et al.* found charge separation in partially strained [110] SiNWs [99]. T. Thonhauser *et al.* studied the phonon modes in [111] SiNWs [100]. Finally, W. X. Zhang *et al.* investigated electron-phonon coupling and transport properties in n-type [110] SiNWs [101].

**4.2 Single Li Insertion in Ultra-thin SiNWs**

Due to confinement effect and surface effect in Si nanowires, distinct Li insertion behavior in SiNWs, in contrast to bulk Si, is expected, and there should be different Li insertion conditions between the surface regions and the inner region in SiNWs. The difference is especially remarkable in ultrathin SiNWs, because as the diameter of the nanowire decreases, quantum confinement is more significant and the surface to volume ratio increases. Therefore, to reveal the influence of these effects, SiNWs with small diameters were used for simulations, and the studies focused on single Li insertion behavior, which is a fundamental step in studying Li insertion in SiNWs. The binding energies and diffusion barriers were chosen as the important computational quantities.

Q. F. Zhang *et al.* investigated single Li atom insertion in various SiNWs with different diameters and along different growth directions to understand the fundamental interactions between Li and SiNWs and also the microscopic process of Li insertion dynamics [91]. Four types of SiNWs with the long axis along the [110], [001], [111], and [112] directions were investigated with diameters ($d$) ranging between 1.0 to 2.5 nm. The cross-sectional planes of these SiNWs are displayed in the left panel of Figure 11. To calculate the binding energies in different Li insertion positions in each type of SiNW, one typical core site (C site, locating in inner region), intermediate site (I site, locating between inner and surface region), and surface site (S site, locating on surface), as marked by red, blue, and green balls, are extracted. The resulting binding energies ($E_b$) are shown in the right panel of Figure 11. Comparing the $E_b$ values, it can be seen that the binding energies are quite distinct in SiNWs with different orientations and diameters, as well as on different insertion sites. The rule of $E_b$ values can be summarized as below: (1) $E_b$ becomes larger when the diameter of the SiNWs increases due to the weakening of quantum confinement, and the series of [110] SiNWs always induce the highest binding energies on various insertion locations, which indicates that [110] SiNWs are more favorably Li-doped. (2) In the same SiNW, $E_b$ for the S site is always the highest, while that on the I site is always the smallest, regardless of the size of SiNWs.

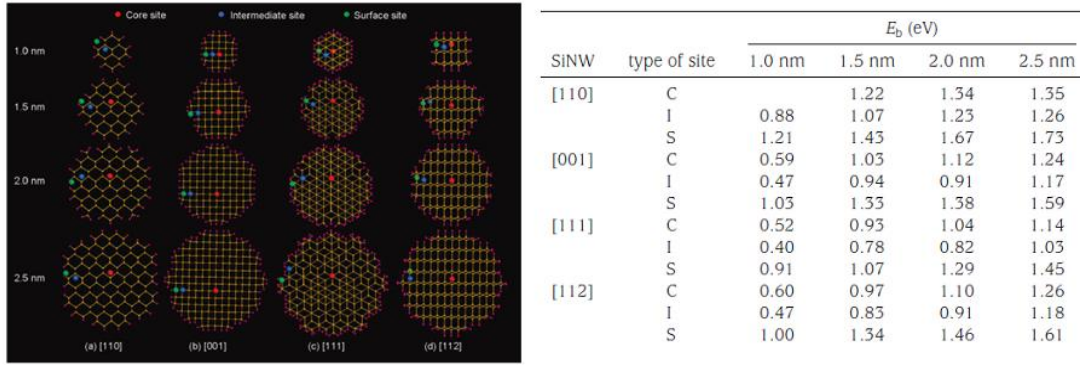

Figure 11. Left panel: The cross planes of SiNWs with axis oriented along (a) [110], (b) [001], (c) [111] and (d) [112], with the diameters $d \approx$ 1.0, 1.5, 2.0 and 2.5 nm. Si and H atoms are represented by small yellow and pink balls, respectively. The typical core, intermediate, and surface sites are shown as larger red, blue, and green balls, respectively. Right panel: Binding Energies ($E_b$) at typical Surface (S), Intermediate (I), and Core (C) Sites in [110], [001], [111] and [112] SiNWs with diameters $d \approx$ 1.0, 1.5, 2.0 and 2.5 nm. The figure is from Ref. [91].

The difference of binding energies can be attributed to the existences of both surface Si and core Si atoms. To investigate the distinction between different Li-Si bonds, the charge transfer was investigated. Figure 12(a) shows the charge transfer along different kinds of Li-Si bonds. The comparison demonstrates that less electron drift occurs from the Li to surface Si than to core Si, which indicates that the interaction between surface Si and Li atoms is weaker than that between core Si and Li atoms. The partial density of states for a surface Si atom ($Si_S$) and a core Si atom ($Si_C$) are shown in Figure 12(b). The sharp peak in the range of -6.8~-6.2 eV indicates very strong interaction between $Si_S$ and the bound H atom, which makes the wavefunction tend to distribute toward the $Si_S$-H bond and weaken the wavefunction overlapping between the $Si_S$ atom and the Li defect. To have a more comprehensive understanding of such surface and passivation effects, the passivated H atom was substituted by a series of halogen atoms or alkali atoms and the energy differences between the I site and C site were computed and listed in Figure 12(c). It can be clearly seen that when the electronegativity (EN) of the substituted passivation element becomes larger, the I site is less stable compared to the C site. This distinction reveals the ionic nature of the Si-Li bond. As the EN of the surface atom increases, the electron density around $Si_S$ decreases and the interaction between Si and Li weakens. Therefore, this suggests that the energy difference between different sites is not mainly determined by the confinement effect, but by the features of the interface atoms. H-passivated SiNWs can also represent the common cases in experimental research, in which elements with high EN values exist on interface, like oxygen or a transition metal [74,102].

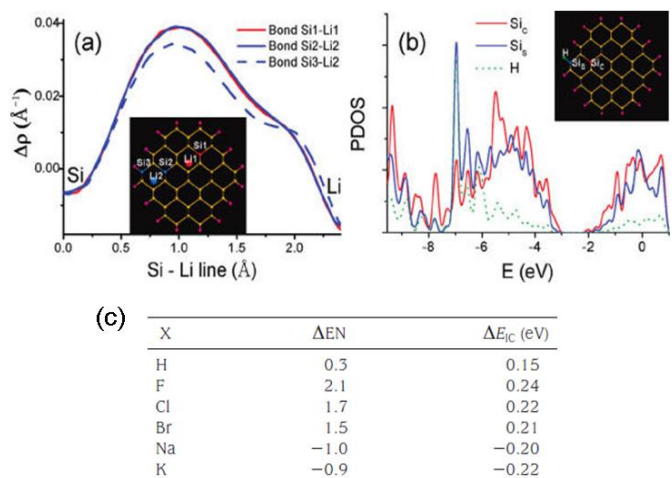

**Figure 12. (a) Charge density difference** $\Delta\rho = \rho[Li/SiNW] - \rho[Li] - \rho[SiNW]$ **along Si-Li bonds. The inset is the [110] SiNW cross-plane, which illustrates the bond Si1-Li1 when Li is on the core (red) site, and bond Si2-Li2 and bond Si3-Li2 when Li is on the intermediate (blue) site. (b) Partial density of states (PDOS) for a surface Si atom ($Si_S$) and a core Si atom ($Si_C$) are shown by blue and red curves, respectively, and the PDOS of H bound to $Si_S$ is also plotted for reference. The inset is the [110] SiNW cross plane, which illustrates the positions of $Si_S$, $Si_C$, and H. (c) Energy difference between the I site and the C site** $\Delta E_{IC} = E_I - E_C$ **when the neighboring surface Si atom for the I site is bound to different types of passivation atoms. The figure is from Ref. [91].**

The barriers for Li atom diffusion were also simulated. Diffusion barriers from the surface to the inner core in [110] and [111] SiNWs with $d \approx 1.5$ nm were extracted. The pathways and corresponding energy curves are shown in Figure 13. The energy variations in both types of SiNWs show a similar behavior. For the diffusion process inside the surface region (from 1 to 3), all the barriers are very low (0.12-0.20 eV), while inside the inner region (from 4 to 7), the barriers are similar to that in bulk Si (0.58 eV). The barriers for the surface-to-inner process (3 to 4) are extremely high, which indicates that it is difficult for a Li defect to diffuse inside from the surface. The result is consistent with the experimental observation that the Li insertion in SiNWs is layer by layer from surface to inner region as the Li concentration increases during the Li insertion process. In another paper, J. Han *et al.* also have a detailed study on single Li diffusion barriers in the core region of SiNWs, and their investigation focused on the small difference of barriers in different regions or diffusion directions [103].

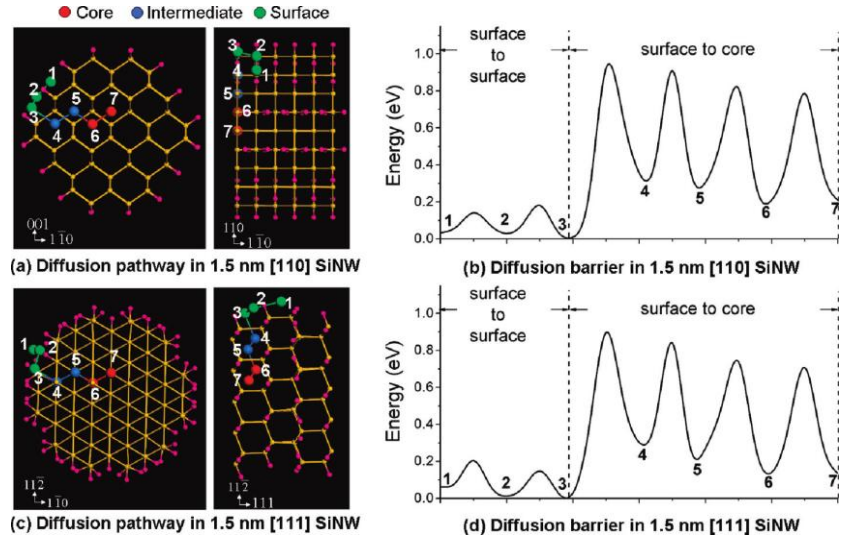

**Figure 13.** Li diffusion in 1.5 nm [110] and [111] SiNWs. (a) and (c) The cross section (left) and side (right) view of the pathways under study. The diffusion barriers along the pathways are shown in panels (b) and (d), respectively. The figure is from Ref. [91].

Imposing external strain is a common way to change the properties of a material and improve device performance, and it is especially important for low-dimensional confined systems. Q. F. Zhang *et al.* studied the anisotropic strain effects in different types of SiNWs when external uniaxial strain is applied to SiNWs, and the investigation concentrated on the variances of binding energies and barriers [104]. The axial length change percentage $\varepsilon = (L - L_0)/L_0$ ($L$ and $L_0$ represent z-axis vector length with and without strain) serves as the parameter to measure the extent of strain, and positive (negative) $\varepsilon$ means that tensile (compressive) strain is applied. Figure 14(a) shows the Li binding energy variances as strain values range from $\varepsilon = -5\%$ to $\varepsilon = 5\%$ in 1.5 nm diameter [110], [001], [111], and [112] SiNWs. It is suggested that the Li binding energy in the [110] SiNW is the most sensitive to strain and depends nearly linearly on strain. The strain effect is smaller in other types of SiNWs, and the [001] SiNWs show the least dependence. At $\varepsilon = 5\%$, $\Delta E_b$ values in all types of SiNWs with diameters of $d \approx 1.5, 2.0, 2.5$ and 3.0 nm were calculated. The $\Delta E_b$ versus $d$ curves, as shown in Figure 14(b), are flat, which indicates that $\Delta E_b$ depends weakly on the SiNW size.

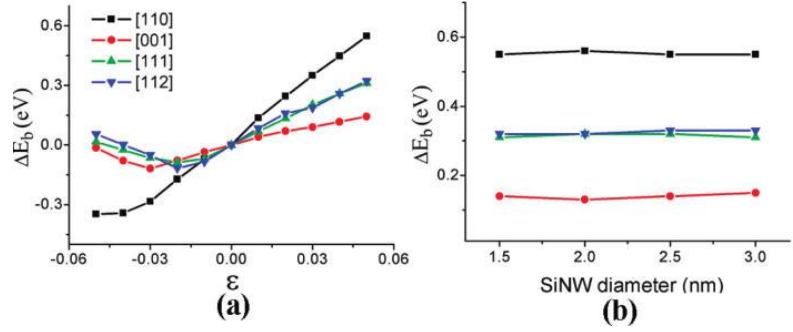

**Figure 14.** (a) Binding energy change $\Delta E_b$ in 1.5 nm [110], [001], [111], and [112] SiNWs from $\varepsilon$ =-5% to $\varepsilon$ =5%. (b) $\Delta E_b$ in four types of SiNWs with diameters $d \approx$ 1.5 and 3.0 nm at $\varepsilon$ = 5%. The figure is from Ref. [104].

Figure 15 shows illustrations of typical diffusion processes and the calculated core-region diffusion barriers $V(\varepsilon)$ in four types of 1.5 nm SiNWs when $\varepsilon$ = -5%, 0% and 5%, which serve as representative compressed, unstrained, and stretched states in the SiNWs. The tendency of barrier variance under strain changes greatly as the diffusion pathway grows parallel to the axis. Under tension (compression), the 1→2 process in [110] or [112] SiNWs, which is perpendicular to the axis, leads to the largest barrier reduction (increase) among all the processes. The 1→3 process in the strained [110] SiNW, for which the axial angle is 35°, has a similar barrier height as an unstrained wire; when the angle between the diffusion pathway and the axis is very small, like in the 1→3 process in [111] or [112] SiNWs, the barrier becomes larger (smaller) than the unstrained case. As a result, the barriers are strongly orientation-dependent in strained SiNWs. That is different from the unstrained SiNWs case, in which the barrier height difference among various diffusion directions is small [91]. Since Li impurities prefer to follow low-barrier diffusion pathways, unbalanced Li diffusion or Li distribution can be expected in strained SiNWs, which suggests promising routes for future electrochemical applications.

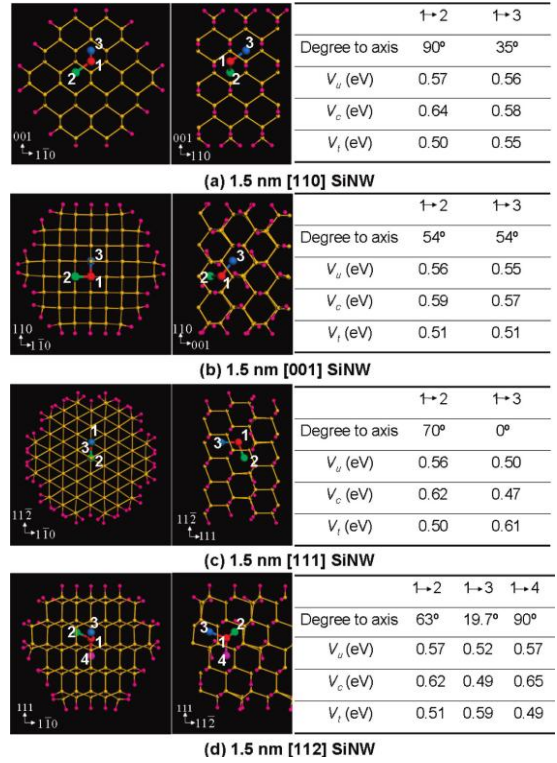

**Figure 15. Illustration of a typical Li diffusion processes in 1.5 nm SiNWs with axial directions along (a) [110], (b) [001], (c) [111], and (d) [112] and their barrier values when $\varepsilon = 5\%$ ($V_c$), 0% ($V_u$), and 5% ($V_t$). The figure is from Ref. [104].**

## 5. Lithiation Dynamics Process at the Li/Si Interface

The insertion of Li into the Si electrode causes great structural changes in the negative electrode. In order to optimize the performance of lithium batteries, it is therefore important to trace these changes during lithiation, the details of which cannot be clearly obtained from experiment [20]. Although some studies have simulated the lithiation process and investigated phase transitions in Li-Si bulk compound systems, as shown in Section 3, they are completely based on static formation energy comparisons, and as a result, the dynamic process and structural evolution during lithiation cannot be modeled. Particularly, remarkable lithiation anisotropy between different crystallographic orientations of Si has been recently observed for microstructures and nanowires, which is a very important phenomenon [105-107]. However, this fascinating effect, which closely relates to dynamic processes on the Si surface or the Li/Si interface, can be completely buried in bulk Li-Si compound models. Therefore, the simulation of lithium insertion, starting dynamically from the Li/Si interface, is quite necessary.

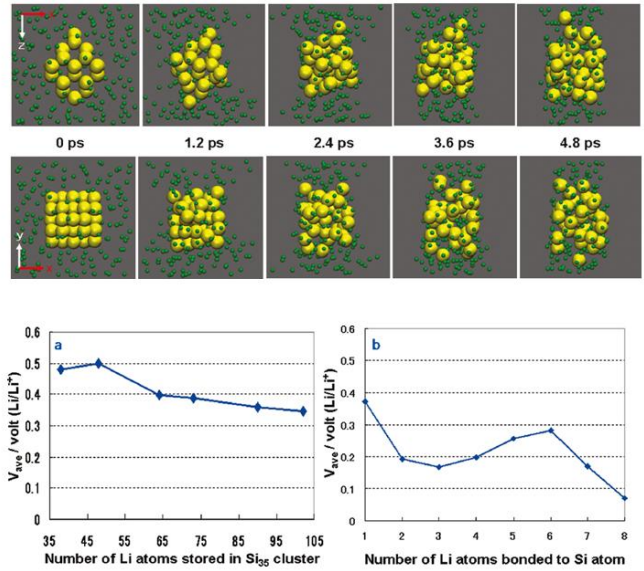

**Figure 16.** Upper panel: Structures at various stages of lithiation of a $Si_{35}$ cluster with 160 Li atoms viewed from the y direction (top) and the z direction (bottom). Yellow and green balls represent Si and Li atoms, respectively. Lower panel: Averaged Li insertion potential of $Si_{35}Li_n$ (n = 38, 48, 64, 73, 90, and 102) cluster (a) and Si atom (b). The figure is from Ref. [108].

Y. Okamoto studied the dynamical lithiation process in a nano-sized silicon cluster containing 35 Si atoms ($Si_{35}$) using an *ab initio* MD scheme [108]. The simulated cell contains a $Si_{35}$ cluster and 160 Li atoms around the cluster in the initial geometry. The configuration evolution during MD dynamics is shown in the upper panel of Figure 16. It is found that the Si nano-cluster expands significantly during lithiation. This is supported by the calculation that during the real-time evolution, the average distance between a Si atom and the center of mass of 35 Si atoms increased from 4.04 (at 0 ps) to 5.50 Å (at 4.8 ps). The averaged Li insertion potential, $V_{ave}$, was examined for six different lithiated cluster geometries of $Si_{35}Li_{38}$, $Si_{35}Li_{48}$, $Si_{35}Li_{64}$, $Si_{35}Li_{73}$, $Si_{35}Li_{90}$, and $Si_{35}Li_{102}$, which were extracted at different lithiation stages. The lower panel of Figure 16 shows $V_{ave}$ of these lithiated clusters. The $V_{ave}$ gradually decreases from 0.48 to 0.34 V as the amount of Li atoms increases, except for $Si_{35}Li_{48}$, which has a somewhat higher potential than that expected from the declining trend. The values of $V_{ave}$ are comparable to the Li insertion potential into bulk Si. For comparison, $V_{ave}$ of a single Si atom instead of $Si_{35}$ was computed. The potential range of the Si atom (0.07 to 0.38 V) is comparable to that of the clusters and bulk. This result suggests that sub-nanosized Si clusters would potentially work as an anode active material for LIBs at least in terms of the insertion potential.

P. Johari *et al.* studied the Li-Si mixing mechanism during lithiation of the Si electrode using *ab initio* MD methodologies [109]. The MD calculations were performed at the high temperature range of 900 ~ 1500 K, and the dynamic lithiation simulations in both crystalline and amorphous Si are carried out. The simulation system contains 64 Si atoms and 64 Li atoms. The starting ($t = 0$ ps), intermediate ($t = 0.3, 0.9, 1.5, 2.1, 3.6$, and $6.0$ ps) and final configurations ($t$ ~15 ps) of atoms during intercalation of Li into c-Si (top panel) and a-Si (lower panel) at 1200 K are shown in Figure 17(a). For the starting atomic models, the Li atoms and Si atoms are completely separated to mimic the stage before lithiation. To analyze the evolution of the structure, the radial distribution functions ($g(r)$) for Li-Li, Li-Si and Si-Si pairs are computed and shown in Figure 17(b). When lithiation occurs, the number of Si-Si and Li-Li neighbors gradually decreases, while the number of Li-Si neighbors gradually increases over time, indicating an evaluation of the Li-Si mixing process. In the case of the c-Si anode, $g_{Si-Si}(r)$ possesses sharp peaks mainly until $t$ ~3.0 ps, while at later instances, the second nearest neighbor peak disappears and the remaining peaks become broader, indicating amorphization. The intensity of the peaks also falls with time and becomes almost constant after 3.0 ps, which suggests the completion of the transformation into the amorphous LiSi phase. For Li-Si mixing beyond 3.0 ps, the $g_{Si-Si}(r), g_{Li-Li}(r)$ and $g_{Li-Si}(r)$ curves for lithiated c-Si show similar features compared with the lithiated a-Si cases, which further confirms the formation of an amorphous structure.

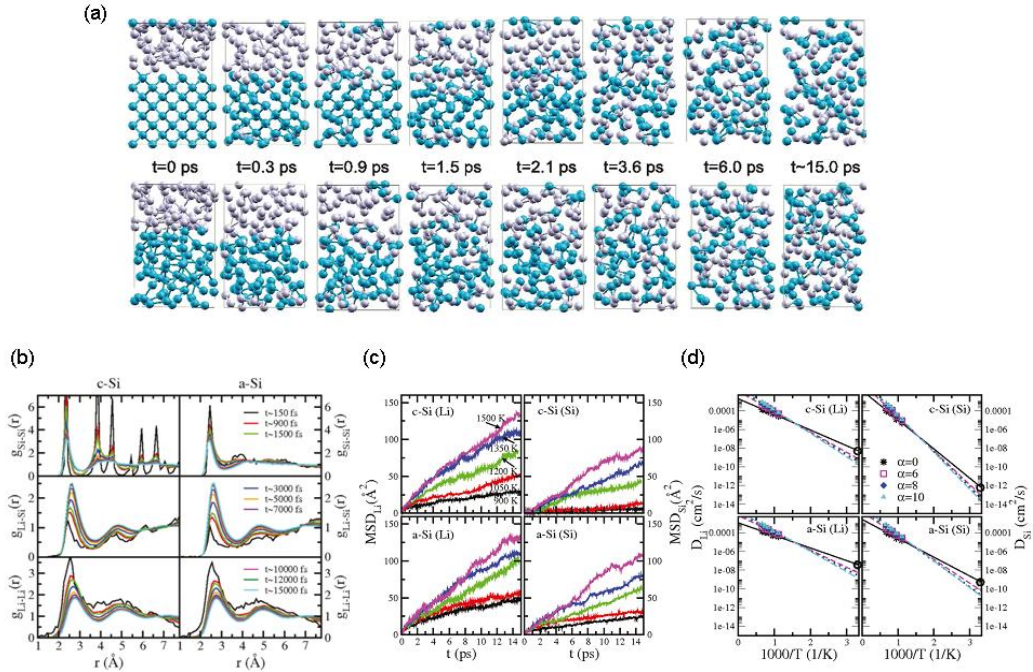

**Figure 17. (a)** Structures at various stages of lithiation of crystalline (top) as well as amorphous (bottom) Si anodes at 1200 K. In the figure, cyan and purple spheres represent Si and Li atoms, respectively. **(b)** The radial pair distribution function $g(r)$ at 1200 K for Si-Si, Li-Si, and Li-Li pairing in c-Si (left) and a-Si (right) anodes at various stages of lithiation, which is represented here in terms of time steps. **(c)** Top graphs present MSD of Li and Si in the c-Si anode while the bottom graphs show it for the a-Si anode with respect to time at 900 K, 1050 K, 1200 K, 1350 K and 1500 K. **(d)** Diffusivity of Li (left) and Si (right) in c-Si (top) as well as a-Si (bottom) with respect to the inverse of temperature. The value of diffusivity is extrapolated to low temperatures using an exponential fit. Diffusivity in the strained cell ($\alpha = 0$) at 300 K is marked by black circle. Diffusivity of Li and Si in unstrained cells is estimated using the formula $D = D_\sigma / \exp(\alpha\sigma/G)$, where $D_\sigma$ is the diffusivity under strain, $\alpha$ is the diffusion transfer constant whose value is chosen in the range of 6 to 10, $\sigma$ is the stress, and $G$ is the modulus of rigidity. Diffusivity at various temperatures, calculated using $\alpha = 6$ (magenta), 8 (blue), and 10 (cyan) are also presented. The figure is from Ref. [109].

In order to determine the diffusivity of Li and Si in Si, the average mean square displacements (MSD) of Li and Si atoms as a function of MD time step for different temperatures are computed. The simulated MSD and diffusivity values are shown in Figure 17(c) and (d), respectively. The diffusivity at room temperature (marked by black circles) is obtained by extrapolating the data at high temperatures. For correcting the effect of artificial stress caused by the high simulated temperature and fixed lattice constant, different diffusion transfer constants are used for calculating diffusivities (see caption of Figure 17 for reference). It is predicted that the diffusivity of Si is much slower than Li in both amorphous and crystalline Si, which suggests that Si atoms are relatively stationary during lithiation. The extrapolated Li diffusivities in c-Si and a-Si at room temperature are in the range of $D_{Li_{c-Si}} = 1.67\times10^{-10} \sim 4.88\times10^{-9}$ cm$^2$ s$^{-1}$ and $D_{Li_{c-Si}} = 1.25\times 10^{-9} \sim 3.69\times10^{-8}$ cm$^2$ s$^{-1}$, respectively. The results lie well within the range of experimentally measured Li diffusivities, from $10^{-14}$ to $10^{-8}$ cm$^2$s$^{-1}$ [110,111].

Recently, M. K. Y. Chan *et al.* reported systemic computational research on lithiation and delithiation processes on various Li-Si interfaces formed on different facets of crystalline silicon, and they simulated a rich range of phenomena observed in experiments, including the amorphization process of Li$_x$Si compounds, crystallization of the fully-lithiated Li$_{15}$Si$_4$ phase and drastic lithiation anisotropy, using a first-principles, history-dependent lithium insertion and removal algorithm (shown in Section 2.2 for reference) [46]. The crystalline silicon surfaces considered in this study are Si(100), Si(110), and Si(111). Figure 18 shows the configurations of the Si (100), (110), and (111) surfaces at various stages of lithium intercalation, and it can be seen that atomistic details of the amorphization process are different for the different surfaces. For the (100) and (110) surfaces, amorphization involves the breakdown into zigzag chains (appearing as dumbbells), with the chains being more ordered (aligned) for (110) than for (100). For the (111) surface, sheets of connected Si are formed before the breakdown into zigzag chains, and some isolated Si atoms are found at high (x≈3−4) Li concentrations. Such difference in breakdown agrees well with the rate of amorphization for wafers with different orientations in experimental measurement [105].

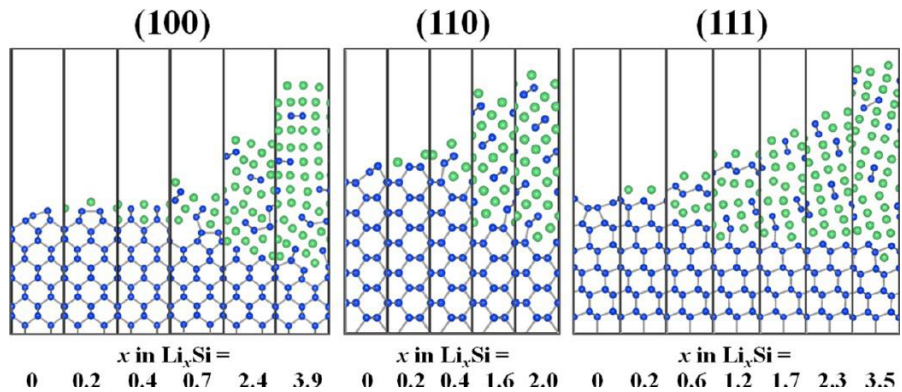

**Figure 18. Configurations of the Si (100), (110), and (111) surfaces at various stages of lithium insertion. The larger (green) spheres represent Li atoms while smaller (blue) spheres represent Si atoms. The bottom of the surface is passivated with H atoms, and the bottom three Si layers are maintained in their bulk equilibrium configurations. The figure is from Ref. [46].**

Figure 19(a) shows the comparison between the calculated voltage curves from the lithiation on Si (100), (110), and (111) surfaces. It can be clearly seen that (110) surface exhibits a higher voltage plateau than the other two surfaces. Using this difference, they qualitatively suggested the reason for drastic lithiation anisotropy observed experimentally is that (110) orientation undergoes dramatic volume expansion [106]. They analyze different kinds of strain energies during lithiation, caused by Si-Si and Li-Li bonds breaking and Si-Li bond formation, and found that insertion of Li is more energetically favorable on (110) surface, and results in higher voltage than that on the (111) and (100) surfaces. Based on this viewpoint, they solved the unsteady-state diffusion equation for a rectangular pillar with the boundary conditions that the side surface has a larger solubility for Li than the top surface, but a scalar diffusion coefficient was used for simulating isotropic diffusion. The time dependence of lithium concentration profiles is illustrated in Figure 19(b). It can be clearly seen that even under isotropic diffusion conditions, the resulting Li concentration profiles demonstrate large anisotropy. The results indicate that the lithiation anisotropy can be explained using only anisotropy in thermodynamics, manifest as a difference in lithiation voltages for different directions, without requiring anisotropic diffusion behavior. Furthermore, they calculated the diffusion barriers for different lithiated Si surface structures, and it was found that the diffusion barriers are somewhat similar, which confirmed that diffusional anisotropy does not play a significant role in the observed lithiation anisotropy.

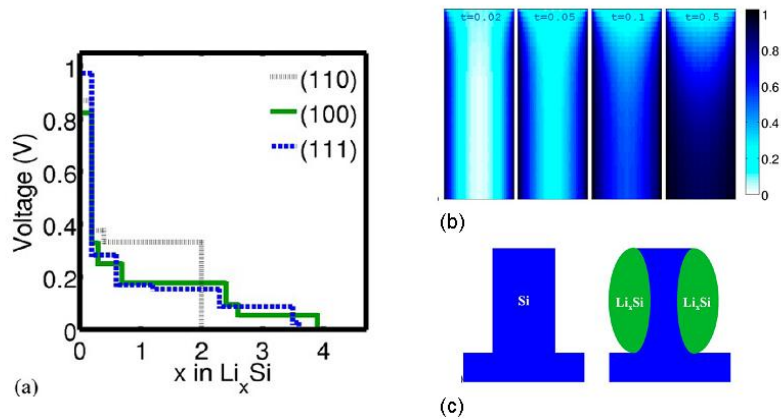

**Figure 19.** (a) Comparison of calculated voltage curves from the lithiation of Si (100), (110), and (111) surface models. (b) Lithium concentration profiles for a microstructured Si column that allows higher solubility of Li on the side wall ((110)) than on the top wall ((111)), as a function of time. The time t is given in units of the diffusion time, $A^2/D$, where $A$ is a characteristic length scale of the structure and $D$ is the diffusion constant. A scalar diffusion coefficient (isotropic diffusion) is assumed. (c) Schematic figure showing the profile of a Si microcolumn before and after lithiation. The figure is from Ref. [46].

The delithiation process, starting from the configuration with the highest Li content, and the relithiation process, starting from the fully delithiated structure, are also simulated using the same algorithm. Comparisons between the formation energies and voltage evolution for the lithiation, delithiation, and relithiation processes are shown as Figure 20. For all three orientations, the average delithiation and relithiation voltages are higher than the average lithiation voltages, consistent with experimental observations [21]. Such hysteresis is caused by the energy difference between the delithiated amorphous Si and the original crystalline Si. It can be expected that amorphous delithiated structures have higher energies than the original crystalline structures; these values are approximately 0.1 to 0.4 eV per Si atom according to the simulation, in reasonable agreement with the measured value of ~ 0.1 eV/atom [112]. There is a much smaller hysteresis between delithiation and relithiation due to amorphization, but the difference of lithiation voltages between the three orientations still exists. Such anisotropy in the second cycle of lithiation suggests that as long as Si micro- and nano-structures are not fully lithiated and a crystalline core is preserved, anisotropy may be preserved in further lithiation cycles.

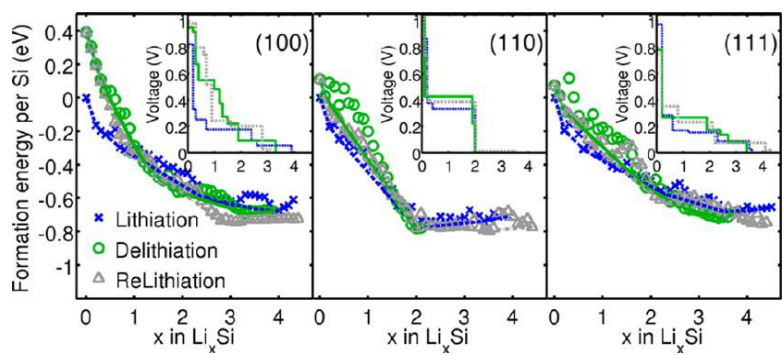

**Figure 20. Comparisons between the convex hulls and voltages for the first lithiation, first delithiation, and second lithiation processes. All Li insertion/removal sites are sampled for the de/re/lithiation process. Only the lowest-energy configuration for each composition is shown. The formation energies are calculated relative to the lowest energy unlithiated Si surface models. The figure is from Ref. [46].**

## 6. Discussion and Perspective

Although much progress has been made, there still exists many unsolved problem in this field. In this section, we will discuss potential future directions, the remining simulation difficulties, and feasible solution methods.

From the reviews shown above, it can be seen that a high percentage of simulation studies concentrate on the study of bulk Li-Si compounds, and various static properties of Li-Si compounds have been clearly investigated. However, progress made in simulating Li insertion into nano-structured Si is still in the initial stage. For Li insertion in SiNWs, studies have modeled steps of single Li atom, which is far from the actual lithiation process for SiNWs in experiments. Furthermore, there has been no attention paid to Li insertion in other kinds of nano-structured Si, like Si nano-cores, nano-shells and nano-tubes, which can have good performance as the lithium battery electrodes as well. The lack of these studies indicates that there still exists many unsolved issues for simulation. Therefore, Li insertion behavior, especially multiple-Li insertion behavior, in different kinds of nano-structured Si, is one of the most important topics for future study. Due to the distinct shape for various Si nano-structures, different Li insertion behaviors are expected. Computational studies on Si nano-crystals provide good opportunities for investigating this issue [113-116]. Among the properties to be studied, the binding energies, diffusion barriers and mechanical properties are the critical points, since they are closely related the electrochemical performance of Si electrodes.

Another important direction is studying the dynamic properties and dynamic lithiation process in Si electrodes, from the fully-unlithiated to the fully-lithiated stage. Although some works have studied this issue, many problems are still unsolved. For example, the anisotropic Li insertion and diffusion on different Si surfaces [106], which is a very important experimental phenomenon, has only been discussed based on empirical models [46], and it has not been directly explained by simulations until now, but the underlying mechanisms still remain unclear. Such facet-dependent behaviors are closely related to the Li/Si and Li/Li$_x$Si interface environments, and can only be completely understood using nano-structured Si system for simulation. SiNWs with different facet orientations can serve as good simulation objects.

Although much understanding can be gained by applying *ab initio* computation, researchers also face many difficulties. The first challenge comes from the great computational cost of simulating nano-structured Si systems. Due to the lack of periodicity in some dimension (two-dimensional periodicity for surface or interface system, one-dimensional periodicity for nano-wire, nano-tube system, and zero-dimensional periodicity for nano-core or nano-shell), there must be numerous atoms for nano-Si configurations. This means that huge computational resources are required for these simulations. For instance, in the simulation of single Li atom insertion in SiNWs, the configurations for ultra-thin SiNWs with diameter of 2.5 nm contain as many as 600~800 Si atoms [91]. For nano-cores or nano-shells, this number will be larger. As the size grows, the number of atoms increases following square-law and cubic-law, respectively. Therefore, it is very tough to treat such large configurations using traditional first-principles methods, even for simulation of static properties like electronic structure computation or geometry optimization.

On the other hand, real-time dynamics properties require step-by-step molecular dynamics methods or newly-developed lithiation schemes, in which the self-consistent electronic structure optimization is the unit step. This means that the amount of calculations dramatically increases, and until now, such simulations can only be performed in bulk or interface systems. Furthermore, the Li atom diffusion barriers in both c-Si and a-Si are as high as 0.5 ~ 0.6 eV, which equals to 6000 ~ 7000 K. This means for the MD scheme, the simulation temperature must be very high in order to overcome the lithiation barrier and room-temperature molecular dynamics is practically impossible, and furthermore, the simulated time must be long enough to achieve the fully lithiated Li-Si phase.

To overcome the difficulties of the large-scale *ab initio* computations mentioned above, it is necessary to apply more advanced methods for electronic structure optimization beyond the conventional plane-wave scheme. The order-*N* scheme is a good choice. The present order-*N* schemes can be classified into two categories. For the first, the linear-scaling computations are performed as the total-energy minimization via a density matrix or a set of Wannier-like localized functions instead of the usual Kohn-Shan eigenfunctions. For the second one, the divide-and-conquer algorithm is used in a local-orbital basis set or in real-space numerical grids. Several mature programs, like SIESTA [117], CONQUEST [118] and ONETEP [119], can be used to perform such simulations. Such simulations have already been performed for some topics that are reviewed above.

Another way to settle this problem is to use tight-binding or empirical-potential models (instead of first-principles schemes) if the computation costs too much even for an order-*N* scheme. In contrast to *ab initio* methods, researchers should introduce reasonable empirical parameters for such simulations, or they may lead to larger error. However, these methods can dramatically reduce the calculation expense, and can efficiently deal with large systems. These methods usually perform well for various pristine Si structures [100,101]. To handle the Li-Si system beyond *ab initio* methods, however, the Li-Si and Li-Li interactions must be accurately depicted.

## 7. Conclusion

From the review above, it can be clearly seen that many efforts have been made in the *ab initio* computational study of lithiation in Si electrodes, including the lithiation process in bulk Si, in Si nano-wires, and at the Li/Si interface. The achievements in this field can be summarized as below:

1. Various properties of crystalline or amorphous bulk Li-Si compounds with different Li concentrations have been studied. The studied $Li_xSi$ compounds cover the whole x range that is observed experimentally. The stable configurations (especially for the crystalline Li-Si compounds) have been clearly identified, and the transition points between crystalline and amorphous phases have been suggested, both in initially-lithiated and fully-lithiated stages. Also, the electron distribution and transfer have been analyzed, and reasonable voltage evolution during the lithiation process has been achieved. Finally, volume expansion has been modeled and elastic properties have been calculating, and elastic softening and plastic deformation behavior during lithiation have been clarified.

2. Electronic properties in Si nanowires and Li insertion behavior in Si nanowires have been studied and are now more fully understood. The studies primarily focus on ultrathin SiNWs and single Li atom insertion. Quantum confinement effects, surface effects and n- or p-doping effects in Si nanowires have been investigated, the binding energy and diffusion barriers for single Li atoms in different type of SiNWs have been calculated, and different kinds of insertion regions have been systematically investigated. In addition, the anisotropic response to external strain for Li insertion has also been studied.

3. The dynamic features of the lithiation and delithiation process have been studied. The investigations focus on the multiple-Li insertion process at the Si/Li interface in order to simulate the actual time-dependent lithiation process in Si electrodes. The insertion of Li atoms into Si nano-clusters has been studied, the detailed Si-Li mixing process and mechanism on the Si/Li interface have been investigated using real-time MD schemes, and lithiation, delithiation and relithiation processes on differently-oriented Si surfaces are simulated and anisotropic diffusion of Li atoms is discussed.

In summary, first-principles simulations have made great progress in the investigation of the microscopic mechanisms of the lithiation process and the atomic-level picture for the Li atom's interaction with various kinds of Si materials. These studies can reasonably explain many lithiation phenomena in experiments, and at the same time, give considerable help and support on optimizing the performance of Si electrode.

## Acknowledgements

The authors thank Dr. M. McDowell for the careful revision of this manuscript. The authors also thank Prof. W. X. Zhang, Dr. W. H. Wan, Dr N. Liu and Dr. G. Zheng for the fruitful discussions and encouragements.

## Reference:

[1] Whittingham M S 2008 *MRS Bull.* **33** 411

[2] Winter M and Besenhard J O 1999 *Electrochim. Acta* **45** 31

[3] Sharma R A and Seefurth R N 1976 *J Electrochem. Soc.* **123** 1763

[4] Boukamp B A, Lesh G C and Huggins R A 1981 *J Electrochem. Soc.* **128** 725

[5] Beaulieu L Y, Hatchard T D, Bonakdarpour A, Fleischauer M D and Dahn J R, 2003 *J. Electrochem. Soc.* **150** A1457

[6] Lee S J, Lee J K, Chung S H, Lee H Y, Lee S M and Baik H K, 2001 *J. Power Sources* **97** 191

[7] Wang W, Kumta P N, 2007 *J. Power Sources* **172** 650

[8] Chan C, Peng H, Liu G, McIlwrath K, Zhang X F, Huggins R A and Cui Y 2008 *Nat. Nanotechnol.* **3** 31

[9] Cui L F, Ruffo R, Chan C K, Peng H and Cui Y 2009 *Nano Lett.* **9** 491

[10] Cui L F, Yang Y, Hsu C M and Cui Y 2009 *Nano Lett.* **9** 3370

[11] Park M H, Kim M G, Joo J, Kim K, Kim J, Ahn S, Cui Y and Cho J 2009 *Nano Lett.* **9** 3844

[12] Kim H, Han B, Choo J and Cho J 2008 *J. Angew. Chem.* **47** 10151

[13] Yao Y, McDowell M T, Ryu I, Wu H, Liu N A, Hu L B, Nix W D and Cui Y 2011 *Nano Lett.* **11** 2949

[14] Hertzberg B, Alexeev A and Yushin G 2010 *J. Am. Chem. Soc.* **132** 8548

[15] Cui L F, Hu L B, Choi J W, Cui Y 2010 *Nano Lett.* **4** 3671
[16] Wu H, Chan G, Choi J W, Ryu I, Yao Y, McDowell M T, Lee S W, Jackson A, Yang Y, Hu L and Cui Y 2012 *Nat. Nanotechnol.* **7** 310
[17] Wu H, Zheng G, Liu N, Carney T J, Yang Y and Cui Y 2012 *Nano Lett.* **12** 904
[18] Hwang T H, Lee Y M, Kong B, Seo J and Choi J W 2012 *Nano Lett.* **12** 802
[19] Li J and Dahn J R 2007 *J. Electrochem. Soc.* **154** A156
[20] Key B, Bhattacharyya R, Morcrette M, Seznéc V, Tarascon J-M and Grey C P 2009 *J. Am. Chem. Soc.* **131** 9239
[21] Key B, Morcrette M, Tarascon J-M and Grey C P 2011 *J. Am. Chem. Soc.* **133** 503
[22] Kang Y-M, Suh S-B and Kim Y-S 2009 *Inorg. Chem.* **48** 11631
[23] Hohenberg P and Kohn W 1964 *Phys. Rev.* **136** B864
[24] Kohn W and Sham L J 1965 *Phys. Rev.* **140** A1133
[25] Perdew J P and Zunger A 1981 *Phys. Rev. B* **23** 5048
[26] Perdew J P and Wang Y 1992 *Phys. Rev. B* **45** 13244
[27] Perdew J P, Burke K and Ernzerhof M 1996 *Phys. Rev. Lett.* **77** 3865
[28] Hamann D R, Schluter M and Chiang C 1979 ¨*Phys. Rev. Lett.* **43** 1494
[29] Vanderbilt D 1990 *Phys. Rev. B* **41** 7892
[30] Blochl P E 1994 ¨*Phys. Rev. B* **50** 17953
[31] Kresse G and Joubert D 1999 *Phys. Rev. B* **59** 1758
[32] Bylander D M, Kleinman L and Lee S 1990 *Phys. Rev. B* **42** 1394
[33] Kresse G and Furthmuller J 1996 *Phys. Rev. B* **54** 11169
[34] Singh D 1989 *Phys. Rev. B* **40** 5428
[35] Segall M D, Lindan P J D, Probert M J, Pickard C J, Hasnip P J, Clark S J and Payne M C 2002 *J. Phys.: Condens. Matter* **14** 2717
[36] Gonze X, Beuken J-M, Caracas R, Detraux F, Fuchs M, Rignanese G-M, Sindic L, Verstraete M, Zerah G, Jollet F, Torrent M, Roy A, Mikami M, Ghosez P, Raty J-Y and D.C. Allan 2002 *Comput. Mater. Sci.* **25** 478
[37] Beck T L 2000 *Rev. Mod. Phys.* **72** 1041
[38] Hirose K, Ono T, Fujimoto Y and Tsukamoto S 2005 *First-Principles Calculations in Real-Space Formalism* (London: Imperial College Press)
[39] Chelikowsky J R 2000 *J. Phys. D: Appl. Phys.* **33** R33
[40] Soler J M, Artacho E, Gale J D, Garcia A, Junquera J, Ordejon P and Sanchez-Portal D 2002 *J. Phys.: Condens. Matter* **14** 2745
[41] Ozaki T and Kino H 2004 *Phys. Rev. B* **69** 195113
[42] Car R and Parrinello M, 1985 *Phys. Rev. Lett.* **55** 2471
[43] Marx D. and Hutter J, 2000 *Ab Initio Molecular Dynamics: Theory and Implementation, in Modern Methods and Algorithms of Quantum Chemistry* pp. 301–449, Editor: J. Grotendorst (John von Neumann Institute for Computing, Forschungszentrum Julich)
[44] Chevrier V, and Dahn J 2009 *J. Electrochem. Soc.* **156** A454
[45] Chevrier V L and Dahn J R 2010 *J. Electrochem. Soc.* **157** A392
[46] Chan M K Y, Wolverton C, and Greeley J P 2012 *J. Am. Chem. Soc.* **134** 14362
[47] Kim H, Chou C-Y, Ekerdt J and Hwang G 2011 *J. Phys. Chem. C* **115** 2514
[48] Aydinol M K, Kohan A F, Ceder G, Cho K and Joannopoulos J 1997 *Phys. Rev. B* **56** 1354.
[49] Mehl M J, Klein R M and Papaconstantopoulos D A 1995 In *Intermetallic Compounds: Principles and Practice;* Westbrook J H and Fleischer R L Eds.; John Wiley and Sons: London Vol. I, Chapter 9, pp 195.

[50] Beckstein O, Klepeis J E, Hart G L W and Pankratov O 2001 *Phys. Rev. B: Condens. Matter* **63** 1341121.

[51] Wan W H, Zhang Q F, Cui Y and Wang E G 2010 *J. Phys.: Condens. Matter* **22** 415501

[52] Mills G and J ónsson H 1994 *Phys. Rev. Lett.* **72** 1124

[53] Jonsson H, Mills G and Jacobsen K W 1998 *Classical and Quantum Dynamics in Condensed Phase Simulations* (Singapore: World Scientific) Chapter 16

[54] Canham L T 1988 *Properties of Silicon* (*Electronic Materials Information Service (EMIS) Datareviews Series No. 4*) Ed Ravi K V, N Hecking, W Fengwei, Z Xiangqin and L N Alexsandrev (London: INSPEC)

[55] Chou C-Y, Kim H, and Hwang G S 2011 *J. Phys. Chem. C* **115** 20018

[56] Zhao K, Wei L. Wang W L, Gregoire J, Pharr M, Suo Z, Vlassak J J and Kaxiras E 2011 *Nano Lett.* **11** 2962

[57] Wen C J and Huggins R A 1981 *J. Solid State Chem.* **37** 271

[58] Kubota Y, Escano M C S, Nakanishi H, and Kasai H 2007 *J. Appl. Phys.* **102** 053704

[59] van Leuken H, de Wijs G A, van der Lugt W and de Groot R A 1996 *Phys. Rev. B* **53** 10599

[60] Kubota Y, Escano M C S, Nakanishi H and Kasai H 2008 *J. Alloys Compd.* **458** 151

[61] Xu Y H, Yin G P, Zuo P J 2008 *Electrochim. Acta.* **54** 341

[62] Chevrier V, Zwanziger J W, Dahn J. R. 2010 *J. Alloys Compd.* **496** 25

[63] Limthongkul P, Jang Y-I, Dudney N J and Chiang Y-M, 2003 *Acta Mater.*, **51**, 1103.

[64] Jung S C and Han Y-K 2012 *Electrochim. Acta.* **62** 73

[65] Nesper R, 1990 *Prog. Solid State Chem.* **20** 1

[66] von Schnering H G, Nesper R and Tebbe K-F, 1980 *J. Curda, Z. Metallkd.* **71** 357

[67] Nesper R, von Schnering H G, 1986 *J. Curda, Chem. Berl.* **119** 3576

[68] Bader R F W, 1990 *Atoms in Molecules –A Quantum Theory*, Oxford University Press, Oxford

[69] Chevrier V L, Zwanziger J W and Dahn J R 2009 *Can. J. Phys.* **87** 625

[70] Beaulieu L Y, Hatchard T D, Bonakdarpour A, Fleischauer M D and Dahn J R 2003 *J. Electrochem. Soc.* **150** A1457

[71] Shenoy V B, Johari P and Qi Y 2010 *J. Power Sources* **195** 6825

[72] Sethuraman V A, Chon M J, Shimshak M, Srinivasan V, Guduru P R 2010 *J. Power Sources* **195** 5062

[73] de Wijs G A, Pastore G, Selloni A and van der Lugt W 1993 *Phys. Rev. B* **48** 13459

[74] Morales A M and Lieber C M 1998 *Science* **279** 208

[75] Holmes J D, Johnston K P, Doty R C and Korgel B A 2000 *Science* **287** 1471

[76] Cui Y, Lauhon L J, Gudiksen M S, Wang J and Lieber C M 2001 *Appl. Phys. Lett.* **78** 2214

[77] Cui Y and Lieber C M 2001 *Science* **291** 851

[78] Duan X, Niu C, Sahi V, Chen J, Parce J W, Empedocles S and Goldman J L 2003 *Nature* **425** 274

[79] Cui Y, Zhong Z H, Wang D L, Wang W U and Lieber C M 2003 *Nano Lett.* **3** 149

[80] Stern E, Klemic J F, Routenberg D A, Wyrembak P N, Turner-Evans D, Hamilton A, LaVan D A, Fahmy T M and Reed M A 2007 *Nature* **445** 519

[81] McAlpine M C, Ahmad H, Wang D W and Heath J R 2007 *Nat. Mater.* **6** 379

[82] Park I Y, Li Z Y, Li X M, Pisano A P and Williams R S 2007 *Biosens. Bioelectron.* **22** 2065

[83] Tian B Z, Zheng X L, Kempa T J, Fang Y, Yu N F, Yu G H, Huang J L and Lieber C M 2007 *Nature* **449** 885

[84] Garnett E C and Yang P D 2008 *J. Am. Chem. Soc.* **130** 9224

[85] Kayes B M, Atwater H A and Lewis N S 2005 *J. Appl. Phys.* **97** 114302

[86] Read A J, Needs R J, Nash K J, L. T. Canham L T, Calcott P D J and Qteish A 1992 *Phys. Rev. Lett.* **69** 1232

[87] Buda F, Kohanoff J, and Parrinello M, 1992 *Phys. Rev. Lett.* **69** 1272

[88] Delley B and Steigmeier E F, 1995 *Appl. Phys. Lett.* **67** 2370

[89] Rurali R and Lorente N 2005 *Phys. Rev. Lett.* **94** 026805

[90] Musin R N and Wang X-Q 2005 *Phys. Rev. B* 71 155318R.

[91] Zhang Q F, Zhang W X, Wan W H, Cui Y and Wang E G 2010 *Nano Lett.* **10** 3243

[92] Ng M-F, Sullivan M B, Tong S W and Wu P 2011 *Nano Lett.* **11** 4794

[93] Lyons D M, Ryan K M, Morris M A and Holmes J D 2002 *Nano Lett.* **2** 811

[94] Leao C R, Fazzio A and da Silva A J R 2007 *Nano Lett.* **7** 1172

[95] Cui Y, Duan X, Hu J and Lieber C 2000 *J. Phys. Chem. B* **104** 5213

[96] Xu X and Servati P 2009 *Nano Lett.* **9** 1999

[97] Han J, Chan T-L, and Chelikowsky J R 2010 *Phys. Rev. B* **82** 153413

[98] Leao C R, Fazzio A and da Silva A J R 2008 *Nano Lett.* **8** 1866

[99] Wu Z, Neaton J B, and Grossman J C 2009 *Nano Lett.* **9** 2418

[100] Thonhauser T. and Mahan G D 2004 *Phys. Rev. B* **69** 075213

[101] Zhang W X, Delerue C, Niquet Y-M, Allan G, and Wang E G 2010 *Phy. Rev. B* **82** 115319

[102] Wu Y, Cui Y, Huynh L, Barrelet C J, David C B and Lieber C M 2004 *Nano Lett.* **4** 433

[103] Chan T-L, and Chelikowsky J R 2010 *Nano Lett.* **10** 821

[104] Zhang Q F, Cui Y and Wang E G 2011 *J. Phys. Chem. C* **115** 9376

[105] Goldman J L, Long B R, Gewirth A A and Nuzzo R G 2011 *Adv. Funct. Mater.* **21** 2412.

[106] Lee S W, McDowell M T, Choi J W and Cui Y 2011 *Nano Lett.* **11** 3034

[107] Liu X H *et al.* 2011 *Nano Lett.* **11** 3312

[108] Okamoto Y 2011 *J. Phys. Chem. C* **115** 25160

[109] Johari P, Qi Y and Shenoy V B 2011 *Nano Lett.* **11** 5494

[110] Xie J, Imanishi N, Zhang T, Hirano A, Takeda Y and Yamamoto O 2010 *Mater. Chem. Phys.* **120** 421

[111] Balke N, Jesse S, Kim Y, Adamczyk L, Tselev A, Ivanov I N, Dudney N J and Kalinin S V 2010 *Nano Lett.* **10** 3420

[112] Donovan E P, Spaepen F, Turnbull D, Poate J M and Jacobson D C 1985 *J. Appl. Phys.* **57** 1795

[113] Melnikov D V and Chelikowsky J R *Phys. Rev. Lett.* **92** 046802

[114] Khoo K H, Zayak A T, Kwak H and Chelikowsky J R *Phys. Rev. Lett.* **105** 115504

[115] Eom J-H, Chan T-L and Chelikowsky J R 2010 *Solid State Commun.* **150** 130

[116] Chan T-L, Tiago M-L, Kaxiras E and Chelikowsky J R 2008 *Nano Lett.***8** 596

[117] Soler J M, Artacho E, Gale J D, Garcia A, Junquera J, Ordejon P and Sanchez-Portal D 2002 *J. Phys.: Condens. Matter* **14** 2745

[118] Bowler D R, Choudhury R, Gillan M J and Miyazaki T 2006 *Phys. Status Solidi b* **243** 989

[119] Skylaris C-K, Haynes P D, Mostofi A A and Payne M C 2006 *Phys. Status Solidi b* **243** 973